\documentclass[11pt]{article}

\usepackage{url}
\usepackage{amsmath}
\usepackage{mathtools}
\usepackage{amssymb}
\usepackage{caption,subcaption}
\usepackage{makecell}

\usepackage[numbers]{natbib}

\usepackage{color}
\providecommand{\blue}[1]{\color{black}{#1}\color{black}\hspace{0pt}}

\definecolor{dred}{RGB}{171,67,53}

\newtheorem{theorem}{Theorem}
\newtheorem{define}[theorem]{Definition}

\def\P{\mathbb{P}}

\def\E{\mathbb{E}}

\def\T{\mathrm{{\scriptstyle T}}}
\def\d{\mathrm{d}}
\def\dt{\mathrm{d}t}
\def\ds{\mathrm{d}s}

\DeclareMathOperator{\He}{Sym}

\DeclareMathOperator*{\diag}{diag}

\def\llongrightarrow{\relbar\joinrel\relbar\joinrel\relbar\joinrel\rightarrow}

\providecommand\X[1]{\boldsymbol{X_{#1}}}
\providecommand\Z[1]{\boldsymbol{Z_{#1}}}
\providecommand\Y{\boldsymbol{Y}}
\providecommand\V{\boldsymbol{V}}
\providecommand\phib{\boldsymbol{\emptyset}}

\providecommand{\rarrow}[1]{\stackrel{#1}{\llongrightarrow}}

\def\Xz{\boldsymbol{X}}

\def\Yz{\boldsymbol{Y}}

\DeclareMathOperator*{\col}{col} 
\usepackage{here}
\usepackage{fullpage}

\providecommand{\bblue}[1]{\color{black}{#1}\color{black}\hspace{0pt}}

\begin{document}

\title{Noise in Biomolecular Systems: Modeling, Analysis, and Control Implications}

\author{Corentin Briat and Mustafa Khammash\\ D-BSSE, ETH-Z\"urich, Switzerland\thanks{\textit{corentin@briat.info; corentin.briat@bsse.ethz.ch}, \textit{mustafa.khammash@bsse.ethz.ch}}}
\date{}

\maketitle

\begin{abstract}
\noindent While noise is generally associated with uncertainties and often has a negative connotation in engineering, living organisms have evolved to adapt to (and even exploit) such uncertainty to ensure the survival of a species or implement certain functions that would have been difficult or even impossible otherwise. In this article, we review the role and impact of noise in systems and synthetic biology, with a particular emphasis on its role in the genetic control of biological systems, an area we refer to as cybergenetics. The main modeling paradigm is that of stochastic reaction networks, whose applicability goes beyond biology, as these networks can represent any population dynamics system, including ecological, epidemiological, and opinion dynamics networks. We review different ways to mathematically represent these systems, and we notably argue that the concept of ergodicity presents a particularly suitable way to characterize their stability. We then discuss noise-induced properties and show that noise can be both an asset and a nuisance in this setting. Finally, we discuss recent results on (stochastic) cybergenetics and explore their relationships to noise. Along the way, we detail the different technical and biological constraints that need to be respected when designing synthetic biological circuits. Finally, we discuss the concepts, problems, and solutions exposed in the article; raise criticisms and concerns about current ideas and approaches; suggest current (open) problems with potential solutions; and provide some ideas for future research directions.\\

\noindent Keywords. \textit{Cybergenetics, Reaction networks, Stochastic Processes, Systems \& Control Theory, Systems \& Synthetic Biology.}
\end{abstract}

\section{Introduction}

\subsection{Gene expression and the central dogma of molecular biology}

DNA is certainly one of the most important substances known to us, for within its sequences lie the genetic instructions for making proteins, the functional workhorses of the living cell. The process of turning the information encoded in DNA into proteins is referred to as gene expression, and it consists of two distinct steps: transcription and translation. Transcription involves copying the genetic information from DNA to messenger RNA (mRNA) using specialized enzymes called RNA polymerases. Translation entails the deployment of complex molecular machines, called ribosomes, to construct cellular proteins using the mRNA as a template. Thus, the flow of sequential information during gene expression generally goes from DNA to RNA to protein. While in some special cases the information can go in reverse (e.g., from RNA to DNA), the central dogma of molecular biology posits that once that information is in the protein, it cannot go back to RNA or DNA.

\subsection{Randomness in biology and life sciences}

It is now well established that noise plays an essential role in natural systems spanning
chemistry, biology, and ecology. For instance, genetic mutations occur randomly, and as
DNA passes from one generation to the next, it accumulates many of these mutations.
Then, natural selection exerts a fitness pressure that favors adapted individuals through
various processes. Thus, the evolution of life can be viewed as the outcome of solving a complex optimization problem in a dynamically changing environment.

While genetic variation is a key mechanism for (random) evolution, it is also fundamental for protecting a population against disease outbreaks or dramatic changes in the environment. So, in the end, variation imparts an evolutionary advantage. At the same
time, at the molecular level cells exploit random noise to achieve certain functions more effectively, as certain key processes, such as gene and chromosomal deactivation, are governed
by highly random events. More generally, the observable characteristics of an individual
(phenotype) are not a deterministic function of the DNA (genotype), as noise in development and other biological processes leads genetically identical individuals to have variations
in their visible characteristics.

Since the seminal works by Adam Arkin, Michael Elowitz, and others \cite{McAdams:97,Elowitz:02,Raser:05}, it has
become well accepted that cells do not necessarily make decisions in a deterministic fashion.
Noise has been observed to be amplified in cells \cite{Hansen:18} (see also \cite{Olsman:18b}) in order to diversify decisions
(e.g., in apoptosis) or produce more variability in phenotypes \cite{Weinberger:05}. In other processes, noise
is filtered \cite{Arias:06} in order to ensure, for instance, fate stabilization \cite{Hansen:18b,Zechner:20}. The study of cellular
noise often focuses on the random cell-to-cell variability in the abundance of the products
of a certain gene expression network that has been observed among genetically identical
cells. It is usually divided into two components, namely intrinsic noise and extrinsic noise.
Intrinsic noise arises due to randomness in the timing and order of chemical reactions at the
molecular scale of the network under consideration. This type of noise cannot be neglected
when molecular copy numbers are low \cite{Elowitz:02} and cannot be fully suppressed by any type of
feedback strategy \cite{Lestas:10}. Extrinsic noise, on the other hand, consists of cell-to-cell variability
due to factors not related to the random nature of gene network under consideration, such
as the metabolic state of the cell, the cell cycle state, and the available cellular machinery,
such as ribosomes and polymerases. Extrinsic noise also arises from fluctuations in the local
cell environment and from the asymmetric partitioning of the molecules after cell divisions
\cite{Chalancon:12}.

\subsection{Homeostasis and perfect adaptation}

An important hallmark of living systems is homeostasis, which is characterized by the
organism’s ability to regulate its internal state (e.g., body temperature and blood glucose
concentration in mammals) despite changing environments. This term was first introduced
by Walter B. Cannon \cite{Cannon:29,Cannon:32}, but the underlying ideas can be traced back to Claude
Bernard and his concept of milieu int\'erieur \cite{Barcroft:31}. Examples of homeostasis can be found
at every level of biological organization (for some examples, see \cite{ElSamad:02,Carroll:16,Schmickl:18,Karsai:20}). At a cellular
level, homeostasis using feedback and feedforward strategies regulates the abundance and
activity of many molecular players. A more stringent type of homeostasis is described by the
concept of perfect adaptation, which is the property of a biological network that enables it to
regulate the quantity of a certain molecule to a fixed value despite the presence of external
stimuli. In control-theoretic terms, this coincides with the regulation problem (with zero
steady-state error) in the presence of constant disturbances.

\subsection{Systems and synthetic biology}

Theoretical analyses of homeostasis and other biological phenomena have now been made
possible through the development of systems biology, which focuses on the development of
tools for the computational and mathematical modeling and analysis of complex biological
systems. Earlier work identified design motifs \cite{Alon:07} that were connected to discovered net-
work properties, such as fold-change detection \cite{Shoval:10}, invariance and equivariance \cite{Shoval:11}, and
so on. The approach is essentially a more holistic (or antireductionist) approach than the more traditional parts-based approach to studying biology. Systems biology, therefore, pro-
vides tools and methods for reverse engineering biological complexity by looking at living
systems as systems that can be analyzed and studied quantitatively (see \cite{Khammash:22IEEE}).

While systems biology presents an approach for reverse engineering, synthetic biology
is best characterized as a discipline for forward engineering biology. It develops the tools
and methods needed to enable the implementation of engineered genetic circuits in order
to modify an existing function or achieve a new desirable function.

At some level, designing synthetic biological circuits is not unlike designing electrical
circuits, a fact that guided the early development of the field. At the same time, there are
substantial differences due to the unique challenges of working with biological substrates
\cite{Ilia:22}. This underscores the need for the development of novel analytical and computational methods for the analysis and design of biological circuits and devices and presents
tremendous opportunities for the further development of the field. Applications of synthetic
biology include the (industrial) bioproduction of molecules ranging from biologics to biofuels \cite{Fortman:08,Galanie:15}, novel solutions for agriculture and environmental remediation \cite{deLorenzo:18}, and medical
therapy and personalized medicine \cite{Kis:15}. The latter include synthetic circuits for personalized medicine such as blood sugar regulation or cancer detection and immunotherapy \cite{Xie:11,Folcher:12}.

\subsection{Cybergenetics}

The development of systems biology and the quantitative analysis of biological systems,
combined with the recent advances in synthetic biology, offers an unprecedented opportunity for control theorists and engineers. Indeed, control theory can help us understand
the cellular mechanisms behind regulation, and in return, biology can provide us with new
control architectures. Moreover, synthetic biology also allows us to implement new controllers in vivo, using biological parts. On the theoretical side, new tools and methods are
needed for the analysis and control of biological systems. The implementation of designed
controllers can then be carried out using synthetic biology and genetic engineering. We refer to this field as cybergenetics \cite{Briat:15e}, which is a portmanteau term that combines Norbert
Wiener’s cybernetics \cite{Wiener:61} and genetics.

Cybergenetics does not aim to be
control theory forced into biology; rather, it is a bottom-up approach that starts from biological observations and suggests concepts and ideas emerging from those observations. It
involves the development of theories, concepts, tools, and methodologies for the modeling,
analysis, and design of biological control systems at the gene level, and hence at the molecular scale. As control systems are fundamental to many biological processes, the potential
applications of cybergenetics are numerous, including industrial biotechnology and medical
therapy, among many other areas \cite{Khammash:22IEEE}.

\blue{\section{Reaction networks}

A reaction network $(\Xz,\mathcal{R})$ consists of a set of $n$ molecular species $\Xz=\{\X{1},\ldots,\X{n}\}$ that interact through $K$ reaction channels $\mathcal{R}=\{\mathcal{R}_1,\ldots,\mathcal{R}_K\}$ denoted as
\begin{equation}\label{eq:RN}
 \mathcal{R}_k:\ \sum_{i=1}^n\zeta_{k,i}^l\X{i}\rarrow{k_i}\sum_{i=1}^n\zeta_{k,i}^r\X{i},\ k=1,\ldots,K
\end{equation}
where $k_i$ is the reaction rate and $\zeta_{k,i}^\ell,\zeta_{k,i}^r\in\mathbb{Z}^n_{\ge0}$ are the left and right stoichiometric vectors. The stoichiometric vector of reaction  $\mathcal{R}_k$ is given by $\zeta_k:=\zeta_k^r-\zeta_k^\ell\in\mathbb{Z}^n$ where $\zeta_k^r=\col(\zeta_{k,1}^r,\ldots,\zeta_{k,n}^r)$ and $\zeta_k^l=\col(\zeta_{k,1}^l,\ldots,\zeta_{k,n}^l)$.
That is, when the reaction $\mathcal{R}_k$ fires, the state moves in the direction $\zeta_k$. The stoichiometry matrix $S\in\mathbb{Z}^{d\times K}$ is defined as $S:=\begin{bmatrix}
  \zeta_1&\ldots&\zeta_K
\end{bmatrix}$. The strength of the reaction $\mathcal{R}_i$ is described by its propensity function $\lambda_i(\cdot)$, which may take different forms depending on the context and the type of kinetics (e.g. mass-action, Hill, Michaelis-Menten, etc.) Finally, it is worth mentioning that reaction networks are multifaceted and can be used to represent any population dynamics such as in ecology, epidemiology, opinion dynamics, multi-agent systems, etc.\cite{Goutsias:13}.}

\blue{\subsection{Deterministic models}

In the deterministic case, reaction networks are quantitatively described in terms of a vector of concentrations, which we denote by $x(t)$, and which evolves on the nonnegative orthant $\mathbb{R}_{\ge0}^n$ or a subset $\mathcal{S}$ of it. The rate of the reaction $\mathcal{R}_i$ is given by the propensity function $\lambda_i:\mathcal{S}\mapsto\mathbb{R}_{\ge0}$ which is defined in such a way that the nonnegative orthant is forward invariant. The dynamical model representing the deterministic reaction network \eqref{eq:RN} is given by the Reaction Rate Equation
\begin{equation}\label{eq:RRE}
  \dot{x}(t)=S\lambda(x(t))=\sum_{k=1}^K\zeta_k\lambda_k(x(t)),\ x(0)=x_0,
\end{equation}
which takes the form of a system of differential equations, meaning that reactions are continuous processes where each reaction $\mathcal{R}_k$ pushes the state in the direction $\zeta_k$ with propensity $\lambda_k(x)$.  Delays and/or spatial effects can be accounted for using delay-differential or partial differential equations.}

\subsection{Stochastic models}

As discussed above,  noise is an essential driving force that impacts all living systems. At the biochemical level, one source of noise arises from the randomness in the firings of the underlying reactions. This randomness can be ignored when the abundance of reactants is high (e.g. in a test tube), but it becomes significant when that some molecules are present in low copy-number, as is often the case in the confines of a single living cell. This then justifies the consideration of stochastic reaction networks.

\blue{\subsubsection{Stochastic reaction networks}

In the stochastic case, reaction networks are quantitatively described in terms of a vector of molecular counts, which we denote by $X(t)$, and which evolves on the nonnegative lattice $\mathbb{Z}_{\ge0}^n$ or a subset $\mathcal{S}$ of it. The propensity functions are now defined in such a way that the nonnegative lattice is forward invariant and thus they may differ from the ones considered in the deterministic setting. In this context, reactions are no longer continuous but instead become discrete and random. In fact, when the reaction $\mathcal{R}_k$ fires, the state jumps from $x$ to $x+\zeta_k$ and the rate of occurrence of those firings is proportionally related to  $\lambda_k(x)$. More formally speaking, under the well-mixedness assumption, which ensures that spatial location of molecules does not play any role in the dynamics, the SRN \eqref{eq:RN} is described by a  CTMC
$(X_1(t),\ldots,X_n(t))_{t\ge0}$ \cite{Anderson:11}, whereby
\begin{equation}\label{eq:Jump}
  \P[X(t+\dt)=x+\zeta_k|X(t)=x]=\lambda_k(X(t))\dt+o(\dt),\ k=1,\ldots,K.
\end{equation}
At the same time, the probability that two different reactions fire at the same time is negligibly small. Notably, this means that the time between two successive firings of the same reaction is exponentially distributed with rate depending on the propensity functions of the reactions. Such networks can be simulated, for instance, using Gillespie's Stochastic Simulation Algorithm (SSA) and any of its many variants; see e.g. \cite{Gillespie:76}.}

\subsubsection{Models}\label{sec:models}

In this section, we discuss the most common ways to describe the evolution of SRNs.\\

\blue{\noindent\textbf{Random time change representation.} The random time change representation simply describes one realization (i.e. a sample-path) of the SRN described by Eq.~\eqref{eq:RN} and Eq.~\eqref{eq:Jump} as
\begin{equation}
  X(t)=X(0)+\sum_{k=1}^K\zeta_kY_k\left(\int_0^t\lambda_k(X(s))\ds\right),
\end{equation}
where the $Y_k$'s are $K$ independent unit-rate Poisson processes. From this expression, one can observe that the trajectory of the state of the process is piecewise constant and will jump at random times (defined by propensity functions) to a new state value depending on the stoichiometric vector of the corresponding reaction. Therefore, the state does not necessarily converge to any specific point in the state-space and will permanently jump from one point to another, unless an absorbing state that traps the trajectory is present. This means that considering a stability concept at the level of the sample-paths is not relevant in general.\\}

\blue{\noindent\textbf{Moment Equations.} Moment equations are an alternative way for representing the evolution of the SRN described by Eq.~\eqref{eq:RN} and Eq.~\eqref{eq:Jump}. The general moment equations for the first- and second-order moments are given by the deterministic system of differential equations
\begin{equation}\label{eq:moments_g}
\hspace{-3mm}\begin{array}{lcl}
    \dfrac{\d \E[X(t)]}{\dt}&=&S\E[\lambda(X(t))],\\
    \dfrac{\d \E[X(t)X(t)^{\T}]}{\dt}&=&\He\{S\E[\lambda(X(t))X(t)^{\T}]\}+S\diag\{\E[\lambda(X(t))]\}S^{\T},
\end{array}
\end{equation}
where $\He[X]=X+X^T$. The rationale for considering such equations is to be able to obtain the moments in a faster way than by using Gillespie's SSA. However, those equations are solvable only in very particular cases when very strong requirements are met by the propensity functions, the stoichiometric vectors and/or the state-space. For most of the interesting cases, those conditions are not met. While nonlinearities often make differential equations analytically not solvable, they can still be numerically integrated. This may not necessarily be the case here as the moment equations may be ``open" in the sense that considered moments (e.g. first- and second-order) will necessarily depend on higher-order ones, making the system of equations ``open". \bblue{A simple example is the network $\phib\rarrow{1}\Xz$, $\Xz+\Xz\rarrow{1}\phib$ for which we have that $\d\E[X(t)]/\dt=1+\E[X(t)]-\E[X(t)^2]$.} This notably prevents us from being able to obtain explicit values for the equilibrium points, by either analytically or numerically solving algebraic equations, and one will have to resort to doing stochastic simulations for computing those equilibrium values, which defeats the purpose for using the moment equations in the first place. To circumvent this, closure methods that aim to substitute those higher-order moments by known, deterministic quantities \cite{Singh:11,Smadbeck:13,Schnoerr:15}, have been proposed.}\\

\blue{\noindent\textbf{Chemical Master Equation.} The Chemical Master Equation (CME), or Forward Kolmogorov's equation, describes the evolution of the probability density function of the CTMC associated with the SRN described by Eq.~\eqref{eq:RN} and Eq.~\eqref{eq:Jump} and is given by
\begin{equation}\label{eq:CME}
  \dfrac{\partial}{\partial t}p(t,x;x_0)=\sum_{k=1}^K\lambda_k(x-\zeta_k)p(t,x-\zeta_k;x_0)-\sum_{k=1}^K\lambda_k(x)p(t,x;x_0),\ x\in\mathcal{S}
\end{equation}
where $p(t,x;x_0):=\P(X(t)=x|X(0)=x_0)$ and $p(0,x;x_0)=\delta(x-x_0)$, where $\delta$ is the Kronecker delta function. The CME is a system of linear differential equations with dimension equal to the cardinality of the state-space $\mathcal{S}$. Similar to the moment equations, the CME is analytically solvable only for very simple networks, and this will not be the case for most interesting networks. Approximate numerical solutions to the CME can be obtained using algorithms such as the Finite-State Projection algorithm \cite{Munsky:08}, which consists of truncating the state-space and integrating forward the truncated system. Perhaps surprisingly, even if in most of the interesting biological applications the CME will consist of a countably infinite system of differential equations, it will be our most essential and useful representation of the dynamics of our SRNs in the context of Cybergenetics. This is further discussed in Section \ref{sec:stability}.}\\

\blue{\noindent\textbf{Approximations.} For completeness, it is important to mention that many approximations of the CTMC exist. Very common ones are the Linear Noise Approximation and the Langevin equation, both of which approximate the dynamics of the CTMC using Stochastic Differential Equations; see e.g. \cite{VanKampen:07}. While those approximations are easier to deal with than the original CTMC, they may fail to capture some of its important properties. Interestingly, it is also possible to show that the deterministic dynamics  \eqref{eq:RRE} coincides, in fact, with the mean-field limit of the CTMC whenever the reaction volume goes to infinity; see e.g. \cite{Anderson:15}. This corresponds to the high-copy-number regime in which molecular concentration is a well-defined quantity. Finally, hybrid models consisting of both deterministic and stochastic dynamics are useful to represent multiscale networks in which molecular species are in both the low and high copy number regimes.}

\section{Noise-induced properties}

While noise or randomness is often seen as a nuisance in engineering applications, it can be sometimes be beneficial in physics and the life sciences, as in the cases of stochastic resonance and coherence-resonance in neuroscience \cite{Roper:98thesis,Gammaitoni:98,Ullner:04thesis}. It is now well accepted that noise may help to achieve a certain dynamical behavior that would be impossible or difficult to achieve in a noise-free setting. In this regard, it seems natural that living organisms may have evolved to exploit that randomness rather than getting rid of it, which would be a much more formidable task. The properties that emerge from, or are facilitated by, the presence of noise are referred to as \emph{noise-induced properties}. In biochemistry, noise-induced properties are those exhibited by an SRN while being absent in the corresponding deterministic reaction network. \bblue{This section emphasizes that extrapolating properties of stochastic networks from their deterministic counterpart, or vice-versa, can be fallacious because their behavior can dramatically differ.}

\subsection{Noise-induced instability and stability}\label{sec:NI:instability}\label{sec:NI:stability}

\begin{figure}
     \centering
     \begin{subfigure}[b]{0.53\textwidth}
         \centering
         \includegraphics[width=\textwidth]{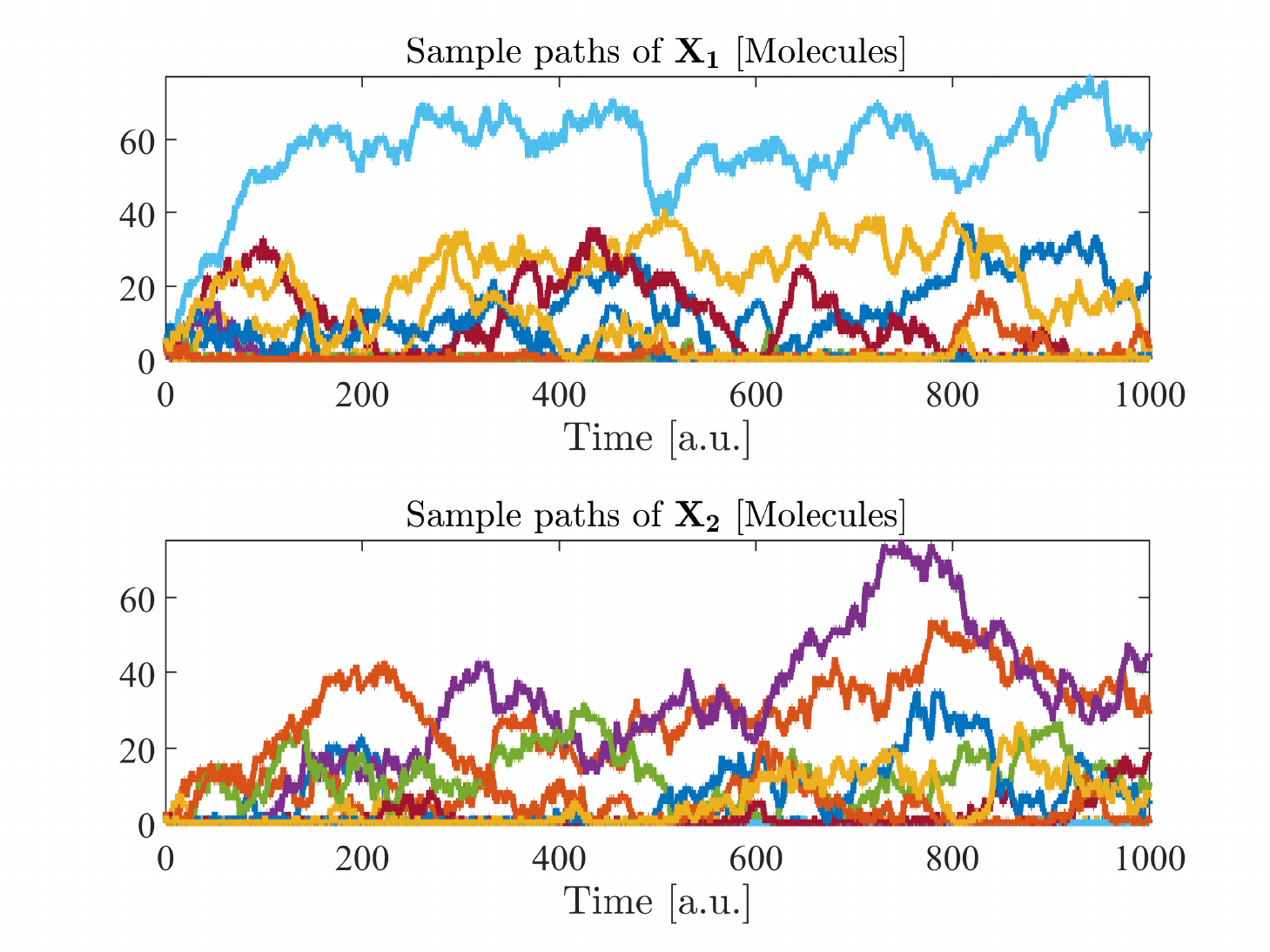}
     \end{subfigure}
     \hfill
     \begin{subfigure}[b]{0.46\textwidth}
         \centering
         \includegraphics[width=\textwidth]{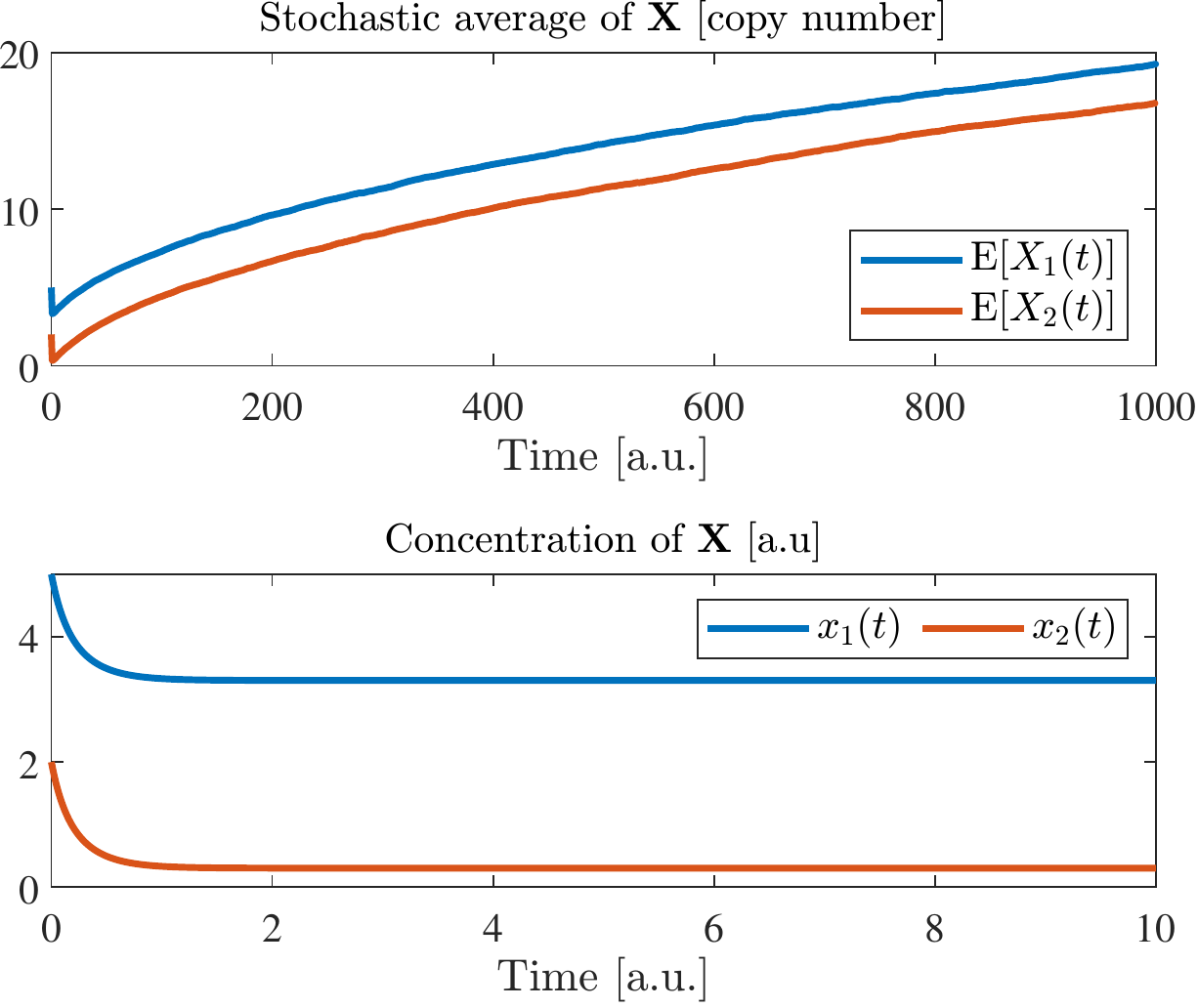}
     \end{subfigure}
     \caption{\textbf{Noise-induced instability.} \textbf{Left.} Five sample-paths of the SRN \eqref{eq:unstable}. \textbf{Top right.} Trajectories of the ensemble average of the states of the SRN. \textbf{Bottom right.} Trajectories of the states of the deterministic network.}\label{fig:unstable}
\end{figure}

\begin{figure}
     \centering
     \begin{subfigure}{0.49\textwidth}
\vfill
\centering\includegraphics[width=\textwidth]{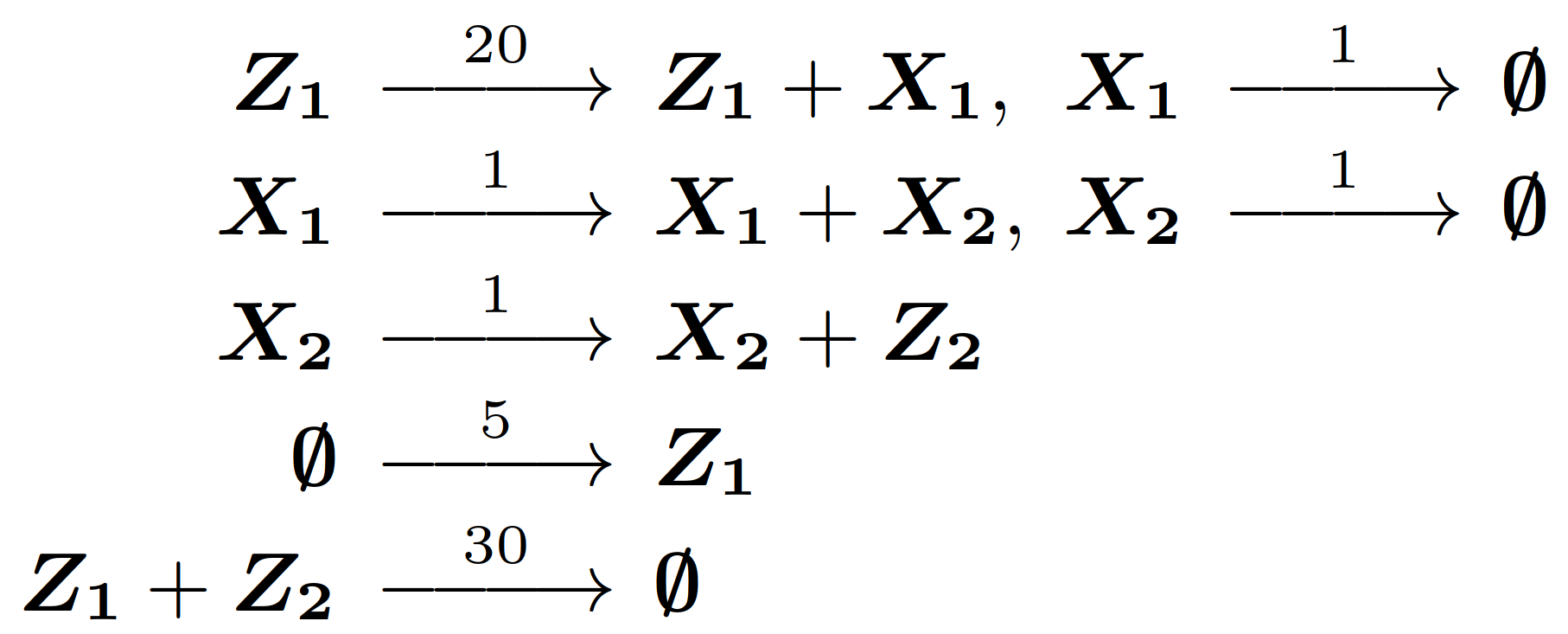}\vspace{15mm}
\vfill
     \end{subfigure}
     \hfill
     \begin{subfigure}{0.49\textwidth}
         \centering \includegraphics[width=\textwidth]{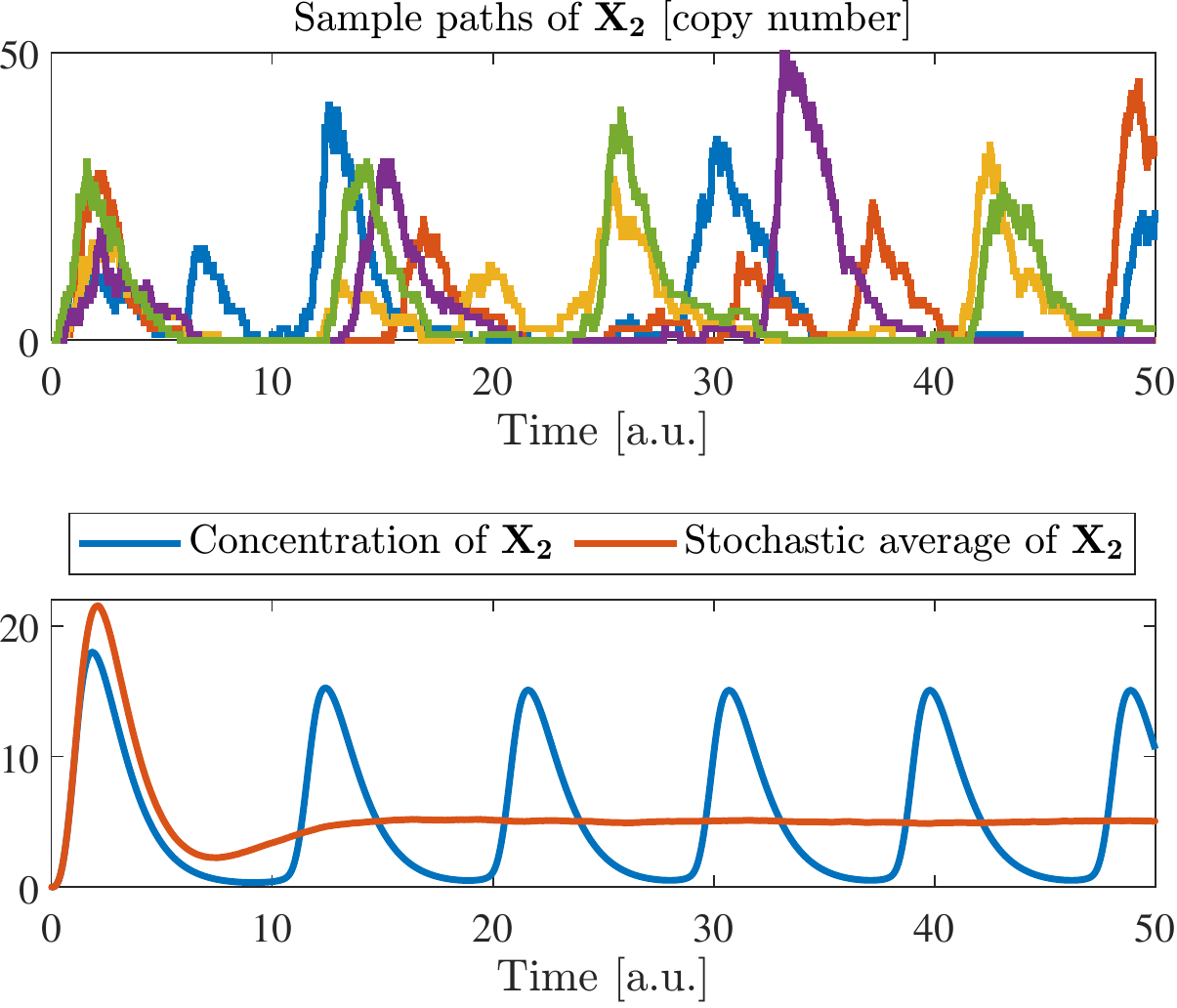}
     \end{subfigure}
     \caption{\textbf{Noise-induced stability.}\textbf{Left.} Reaction network. \textbf{Top right.} Five sample-paths of the species $\X{2}$. \textbf{Bottom right.} Deterministic trajectory (blue) and mean trajectory (red) for the species $\X{2}$.}\label{fig:stable}
\end{figure}

The first noise-induced property, which is also a caveat, is the possible loss of stability due to the presence of noise. Consider, for instance, the following network \cite{Briat:13i}
\begin{equation}\label{eq:unstable}
  \phib\rarrow{1}\X{1},\ \phib\rarrow{1}\X{2},\ \X{1}+\X{2}\rarrow{1}\phib.
\end{equation}
While the unique equilibrium point for the deterministic dynamics is globally exponentially stable, the stochastic dynamics are unstable and all the moments of the CTMC diverge, as illustrated in Fig.~\ref{fig:unstable}. This shows that, in general, one cannot infer stability of the stochastic dynamics from that of the deterministic dynamics. Conversely, noise may also lead to moment convergence in the stochastic setting while deterministic dynamics exhibits an oscillatory behavior as illustrated in Fig.~\ref{fig:stable}. In Section \ref{sec:cybergenetics}, we show that this noise-induced property plays an essential role in the control of SRNs.

\subsection{Noise-induced switching}\label{sec:NI:switching}

The presence of noise allows certain circuits to switch from one stable position to another in a spontaneous way -- a phenomenon also called stochastic resonance -- while the deterministic dynamics converges to one of the stable equilibrium points. A notable example is the stochastic switch of Gardner et al. \cite{Gardner:00} described by $\phib\rarrow{f_1(X_2)}\X{1}$, $\phib\rarrow{f_2(X_1)}\X{2}$, $\X{i}\rarrow{1}\phib$, for $i=1,2$, whose deterministic dynamics exhibits two stable equilibria and one unstable one, and cannot spontaneously switch between them. \bblue{The behavior of the stochastic network is illustrated in Fig.~\ref{fig:switch_focusing} for the propensity functions $f_1(X_2)=50/(1+X_2^{2.5})$ and $f_1(X_2)=16/(1+X_1)$. The deterministic dynamics converges, however, to one of the two stable equilibrium points depending on the initial condition and, therefore, does not exhibit any switching behavior.}

\subsection{Noise-induced amplification}\label{sec:NI:amplification}

\begin{figure}
     \centering
     \begin{subfigure}[b]{0.49\textwidth}
         \centering
         \includegraphics[width=\textwidth]{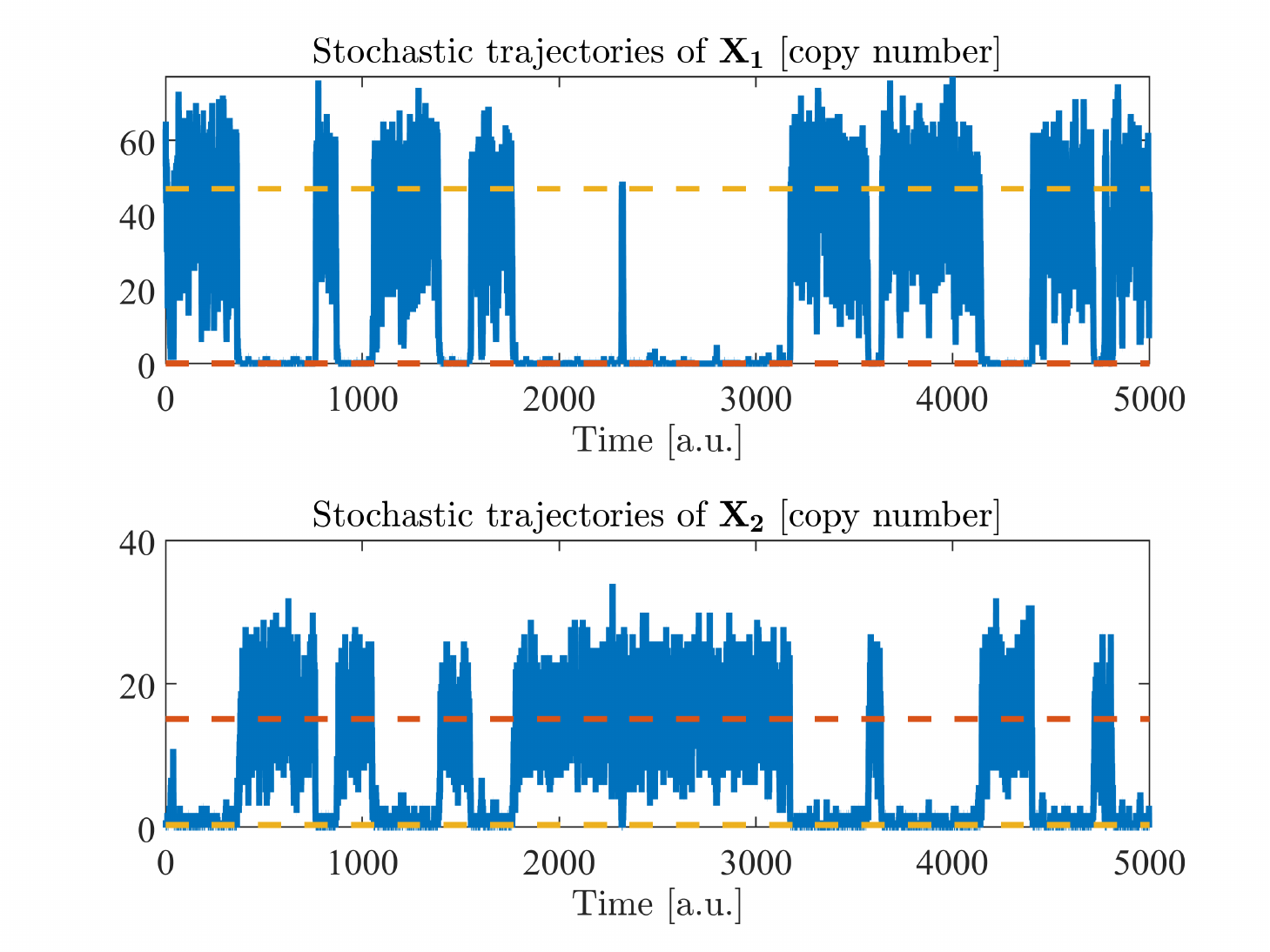}
     \end{subfigure}
     \hfill
     \begin{subfigure}[b]{0.49\textwidth}
         \centering
         \includegraphics[width=\textwidth]{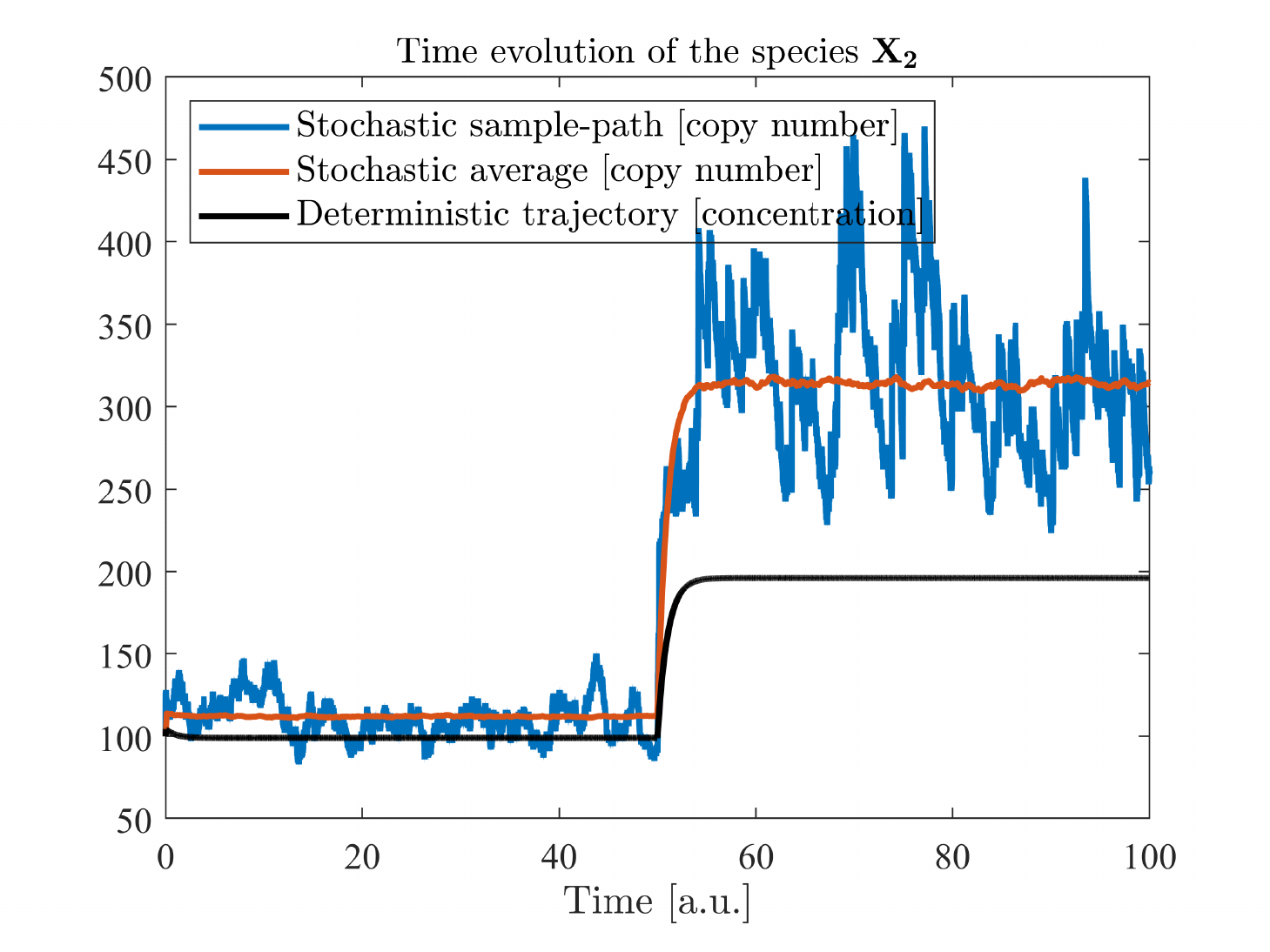}
     \end{subfigure}
     \caption{\textbf{Left: Noise-induced switching.} Evolution of a sample-path of the stochastic switch. The two horizontal lines denotes the two stable equilibrium points of the deterministic network. \textbf{Right: Noise-induced amplification.} Comparison between the deterministic trajectory, a sample-path of the stochastic network, and the average of sample-paths. One can observe that the stochastic trajectories reach much higher values than those in the deterministic case.}\label{fig:switch_focusing}
\end{figure}

Noise-induced amplification, also known as stochastic focusing, was first pointed out by Paulsson et al. \cite{Paulsson:00} and further studied by Khammash and colleagues \cite{Milias:15,Gupta:17b}. This is the phenomenon where the sample-paths and the first-order moments of a particular molecular species take larger values than their deterministic counterparts. A simple example of a network exhibiting such a behavior is the following simplified enzymatic network $\phib\rarrow{k_s}\X{1}\rarrow{100X_1}\phib$, $\phib\rarrow{q(X_1)}\X{2}\rarrow{X_2}\phib$ whose behavior is depicted in Figure~\ref{fig:switch_focusing} for the parameters $q(X_1)=10^4/(1+X_1/0.1)$ and $k_s=1000$, initially. At time $t=45$, $k_s$ is divided by 2.

\subsection{Noise-induced oscillations}\label{sec:NI:oscillations}

\begin{figure}
       \centering
     \begin{subfigure}[b]{0.35\textwidth}
         \centering
         \includegraphics[width=\textwidth]{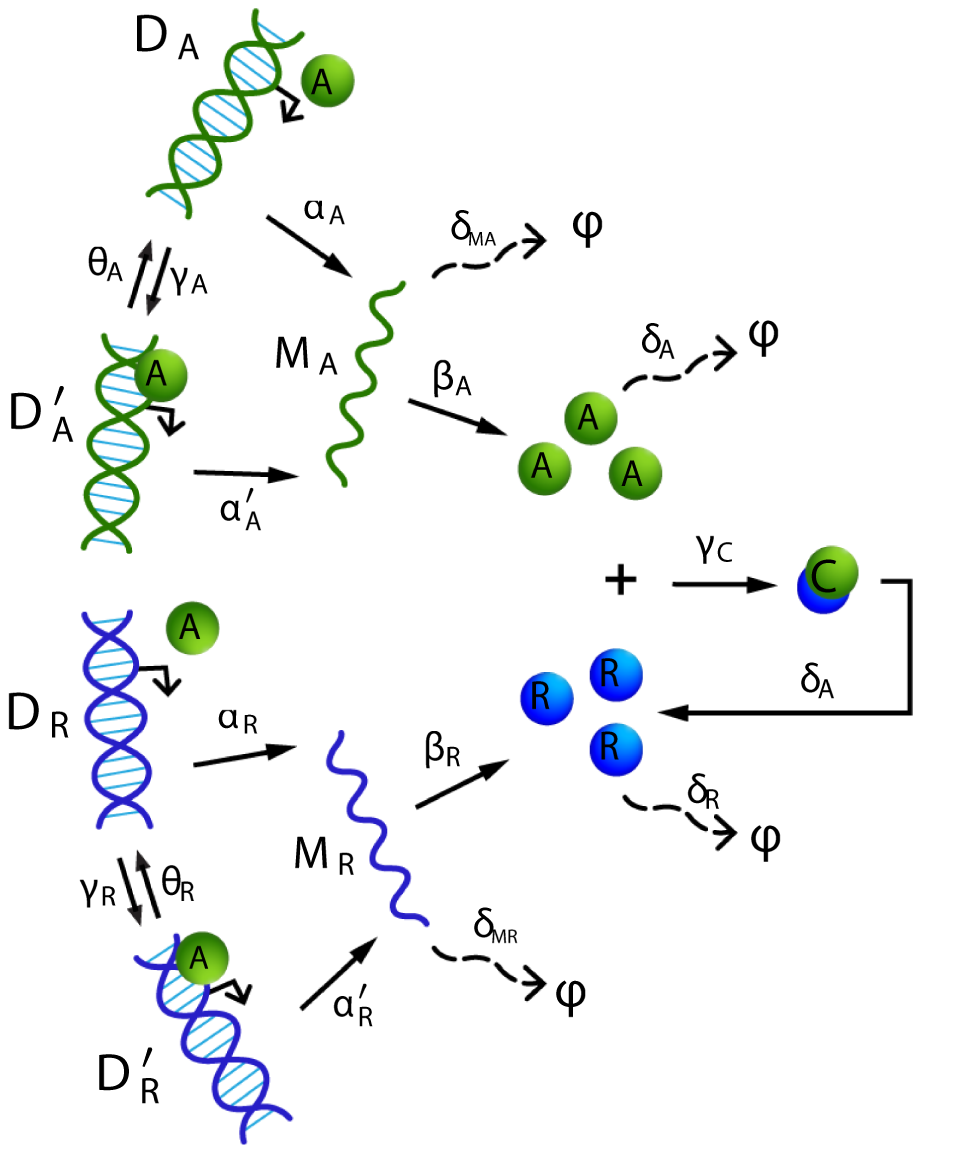}
     \end{subfigure}
     \hfill
     \begin{subfigure}[b]{0.55\textwidth}
         \centering
         \includegraphics[width=\textwidth]{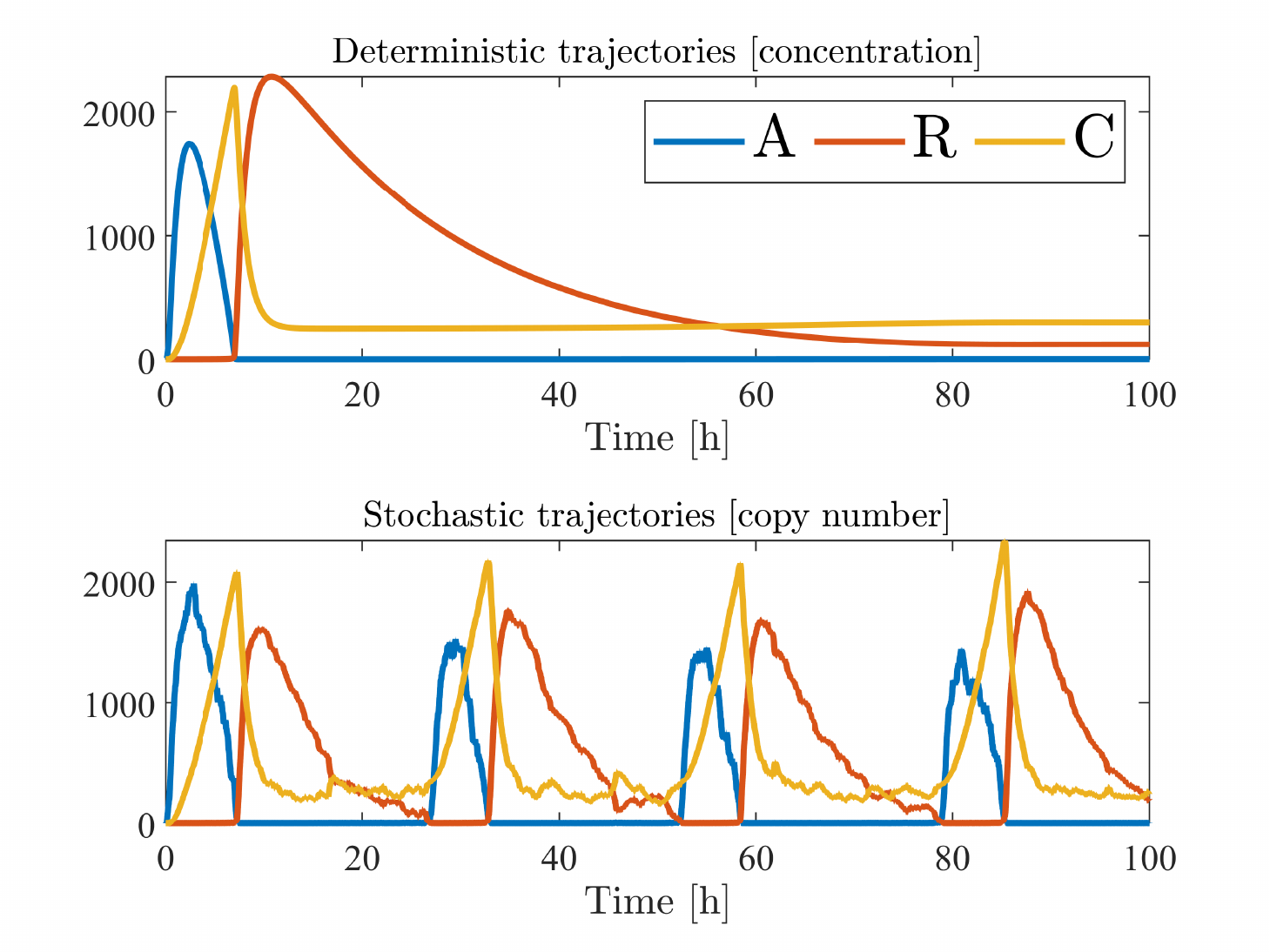}

     \end{subfigure}
     \caption{\textbf{Noise-induced oscillations.} \textbf{Left.} Circadian clock-model of \cite{Vilar:02}. Trajectory of the repressor molecule $\mathbf{R}$ in the deterministic setting (\textbf{Top right}) and in the stochastic setting (\textbf{Bottom right}) .}\label{fig:circadian}
\end{figure}

\blue{In contrast to the noise-induced stability phenomenon is the phenomenon of noise-induced oscillations, which allows certain stochastic networks to maintain an oscillatory behavior for a wider range of parameter values than in their deterministic counterparts. Noise makes the oscillatory behavior more robust. An example of this is the circadian clock network of Vilar et al. \cite{Vilar:02} as illustrated in Fig.~\ref{fig:circadian}.}

\subsection{Noise-induced entrainment}\label{sec:NI:entrainment}

\blue{Entrainment refers to the phenomenon in which an oscillating system with a certain natural frequency $\omega_0$ is driven to another frequency $\omega_1$ through the action of an input signal with frequency $\omega_1$. A well-known example of entrainment in biology is that of the circadian clock, which is entrained by external day-night cycles. Entrainment occurs only if the input signal's amplitude is above a certain threshold which increases monotonically with the distance between the natural and the input frequencies, giving rise to a V-shaped entrainment region (called the Arnol'd tongue) in the (input) amplitude-frequency space.  Gupta et al. \cite{Gupta:16} recently discovered that noise facilitates entrainment at the population-averaged level by expanding the Arnol'd tongues (see Figure~\ref{fig:entrainment}) for the circadian clock model of Leloup \& Goldbeter \cite{Leloup:03}.

\begin{figure}
       \centering
       \includegraphics[width=\textwidth]{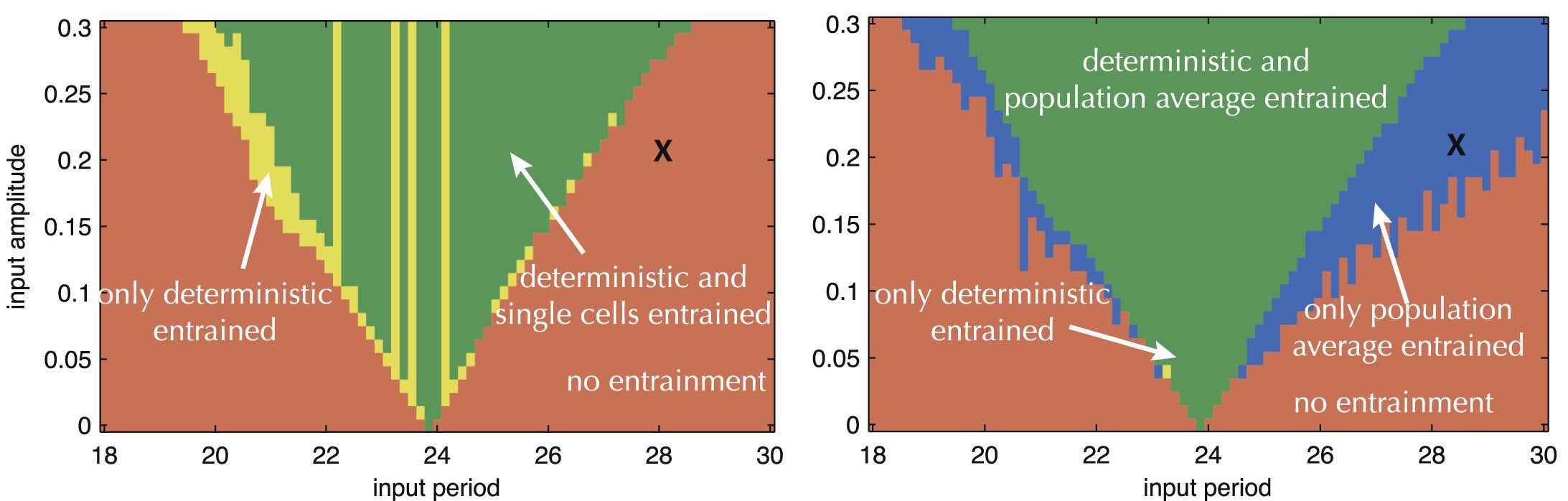}
     \caption{\textbf{Noise-induced entrainment.} Arnol'd Tongues comparing the regions of entrainment between the deterministic and the stochastic settings for the circadian clock model considered in \cite{Leloup:03,Gupta:16}. For the latter, single-cell dynamics is considered on the left while population-averaged dynamics is considered on the right. Observe that for the population-averaged case, the Arnol'd Tongues are significantly expanded in comparison to the deterministic scenario, while this is not the case for the single-cell dynamics.}\label{fig:entrainment}
\end{figure}}

\subsection{Finite-time absorption}\label{sec:NI:extinction}

\begin{figure}
     \centering
     \begin{subfigure}[b]{0.49\textwidth}
         \centering
         \includegraphics[width=\textwidth]{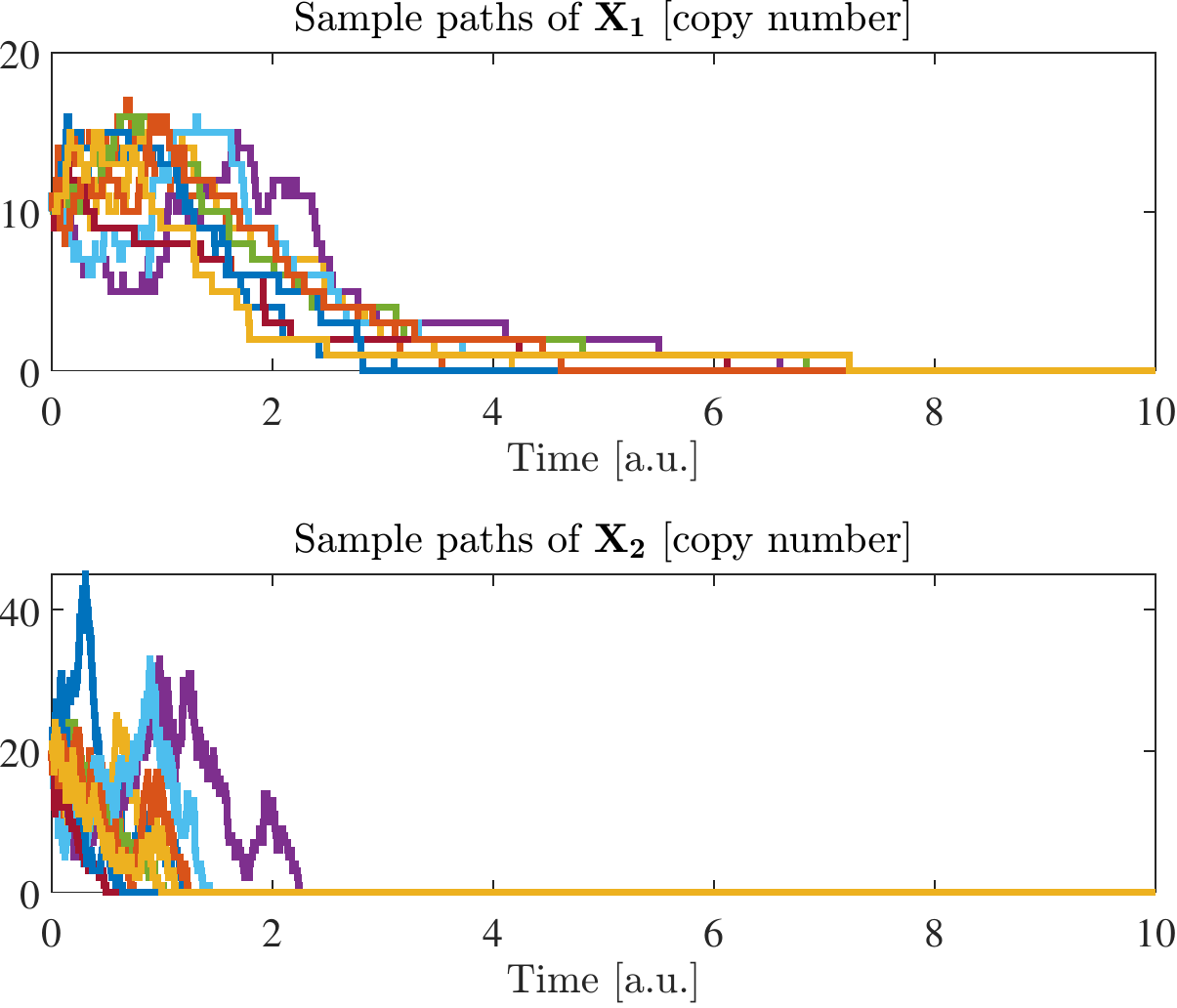}
     \end{subfigure}
     \hfill
     \begin{subfigure}[b]{0.49\textwidth}
         \centering
         \includegraphics[width=\textwidth]{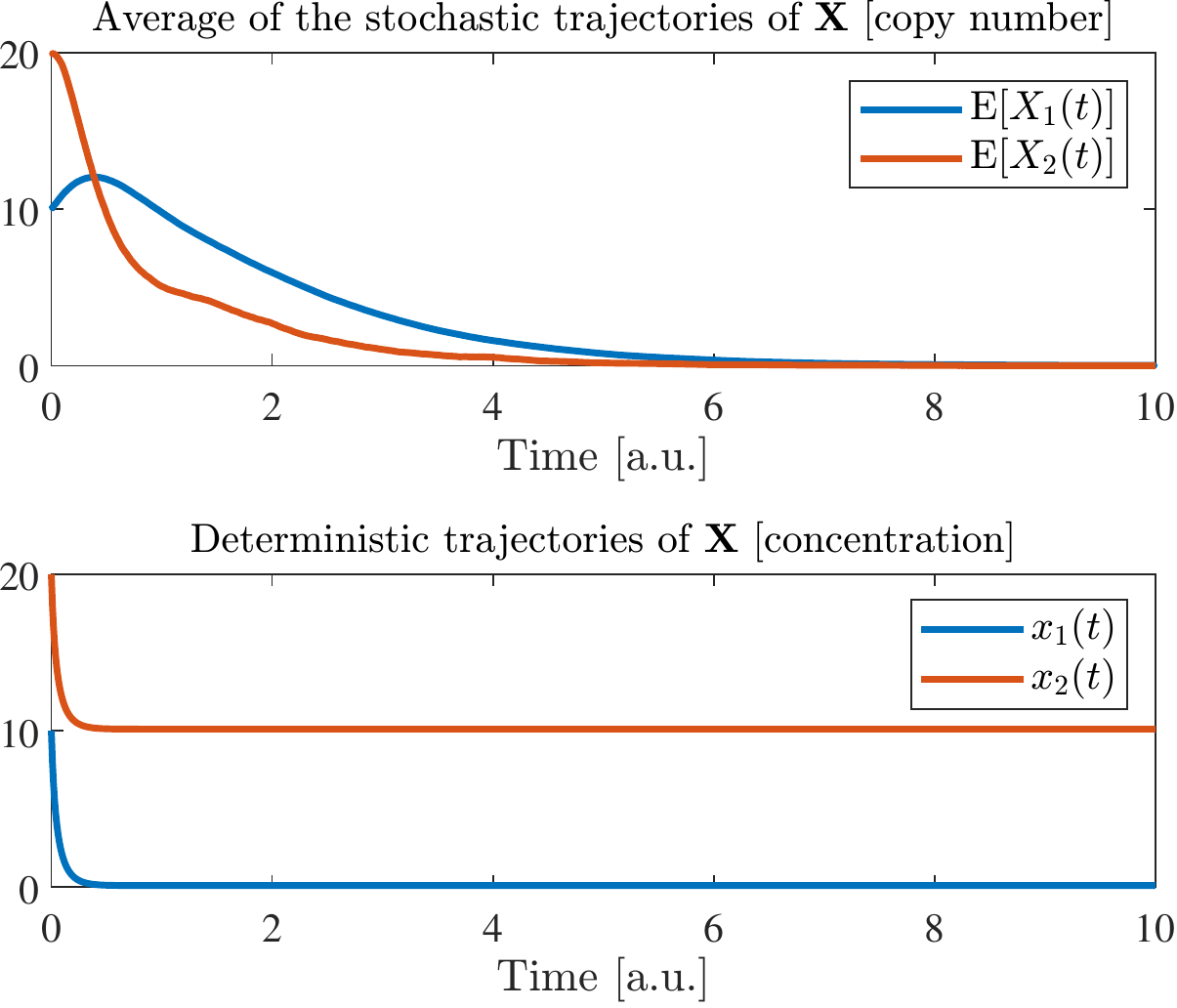}
     \end{subfigure}
     \caption{\textbf{Finite-time absorption.} \textbf{Left.} Sample-paths of the SRN. \textbf{Top right.} Mean of the SRN. \textbf{Bottom right.} Deterministic trajectories.}\label{fig:extinction}
\end{figure}

State trajectories of deterministic networks are attracted by stable equilibria and repelled by unstable ones. However, when stochastic networks are considered instead, it is possible that the state trajectories reach state values corresponding to the unstable equilibrium points of the deterministic dynamics and become trapped there. State values for which the propensity functions are all zero are called \textbf{absorbing states}, and their presence often results in undesirable behavior for the SRN. Importantly, when absorbing states exist and are reachable, they will be reached in finite time with probability one, meaning that the trajectory will necessarily be ultimately trapped there. To illustrate this, consider the network $\X{2}\rarrow{a_2}\X{1}+\X{2}$, ${\X{1}\rarrow{a_1}\phib}$, $\X{2}\rarrow{b_1}2\X{2}$, $\X{1}+\X{2}\rarrow{b_2}\X{1}$. The equilibrium point $(0,0)$ is unstable for the deterministic dynamics whereas the point $(b_1/b_2,a_1b_1/(a_2b_2))$ is globally asymptotically stable. However, for the stochastic dynamics, $(0,0)$ is an absorbing state and the state trajectories will converge to it in finite time with probability one, as illustrated in Figure~\ref{fig:extinction}. 


%

\section{Ergodicity analysis of stochastic reaction networks}\label{sec:stability}

\blue{The first step toward devising circuits and algorithms that can provably solve cybergenetic problems for noisy systems, such as the filtering and the control of SRNs, is the consideration of concepts and the development of tools characterizing stability properties of SRNs. In this section, we argue that the concept of ergodicity is a proper way to approach that problem and discuss its implications.}

\blue{\subsection{State-space irreducibility}

The irreducibility of a state-space $\mathcal{S}\subseteq\mathbb{Z}_{\ge0}^n$ of an SRN is an important property that characterizes the behavior of that network on that space. A state-space is said to be \textbf{irreducible} if any state in it can be reached from any other state in it for the dynamics of the SRN with a positive probability in finite time. An irreducible state-space in the finite state space case consists of all states that are positive-recurrent for the underlying Markov chain \cite{Norris:98}. To verify the irreducibility property of a state-space of an SRN, with either a finite or an infinite state-space, it is enough to show that every state in it can be reached from any other state through a sequence of reactions with positive propensities, which may be cumbersome to check for large networks. Gupta \& Khammash \cite{Gupta:18} proposed a systematic approach based on some ideas from Paulev\'e et al. \cite{Pauleve:14} that were developed for deterministic discrete-time reaction networks.}

\subsection{Definitions and properties of ergodicity}

\blue{Let us start with the definition of ergodicity:
\begin{define}
  The SRN described by Eq.~\eqref{eq:RN} and Eq.~\eqref{eq:Jump} is said to be ergodic if there exists a probability distribution $\pi$ on $\mathcal{S}$ such that $p(t,\cdot;x_0)\to \pi(\cdot)$ as $t\to\infty$ for all $x_0\in\mathcal{S}$.
\end{define}
In order words, the ergodicity property of an SRN coincides with the existence of a globally attracting fixed point for the CME.  Ergodicity implies that for any real-valued function $f$ satisfying $\E_\pi[|f|]<\infty$, where $\E_\pi$ denotes expectation at stationarity, we have that
\begin{equation}\label{eq:ensenble:erg}
  \lim_{t\to\infty}\E[f(X(t))]=\E_\pi[f(X)],\ \textnormal{for all } X(0)\in\mathcal{S}.
\end{equation}
Moreover, the following limit holds with probability 1
\begin{equation}\label{eq:single:erg}
  \lim_{t\to\infty}\dfrac{1}{t}\int_0^t f(X(s))\ds=\E_\pi[f(X)],\ \textnormal{for all } X(0)\in\mathcal{S}.
\end{equation}
The above expressions demonstrate the relevance of the ergodicity concept in the context of SRNs and their applications. To clarify this in the biological setting, consider a cell population where every cell contains the same SRN with state $X$ and let $f(X)$ be some observable. Then, the quantity $\E[f(X(t))]$ denotes the average value of that observable across the cell population at time $t$ whereas $\frac{1}{t}\int_0^t f(X(s))\ds$ is the time-average until time $t$ of that observable over one realization of the SRN, that is, the time average over one single cell trajectory. Equation \eqref{eq:ensenble:erg}  can be used to show that the moments of the reaction dynamics converge to their steady-state values as $t\to\infty$, whereas \eqref{eq:single:erg} is just the ergodic theorem for Markov processes \cite{Norris:98} and shows that the stationary distribution of the population can be inferred by observing a single trajectory of the underlying CTMC for a suﬃciently long time. Therefore, the ergodicity property introduces a \textbf{single-cell/population correspondence} from which we can draw conclusions about the behavior of a single cell from the observation of a cell population and vice-versa.}

\subsection{Establishing ergodicity}

While the ergodicity of a CTMC with finite state-space directly follows from the irreducibility property of its state-space, the case of an infinite state-space requires additional care. \bblue{This additional care can be carried out by the direct analysis of the behavior of the stochastic process (e.g. moment boundedness) or through the consideration of a more general, implicit approach relying on the use of \textbf{Foster-Lyapunov Functions} \cite{Meyn:93}, which are analogues of Lyapunov functions in the current setting.} Among the many results obtained by Meyn \& Tweedie \cite{Meyn:93}, we focus here on the exponential ergodicity result which we also specialize to the case of SRNs:
\begin{theorem}[\cite{Meyn:93,Briat:13i}]\label{th:ergodicity}
The SRN described by Eq.~\eqref{eq:RN} and Eq.~\eqref{eq:Jump} with state-space $\mathcal{S}$ is exponentially ergodic if
\begin{enumerate}
  \item its state-space $\mathcal{S}$ is irreducible, and
  \item there exists a norm-like function $V:\mathcal{S}\mapsto\mathbb{R}_{\ge0}$ (i.e. it is radially unbounded) such that
  \begin{equation}
  \sum_{k=1}^K\lambda_k(x)[V(x+\zeta_k)-V(x)]\le -c V(x)+d
\end{equation}
holds for some $c,d>0$ and all $x\in\mathcal{S}$.
\end{enumerate}
Moreover, in such a case, there exists a unique probability distribution $\pi$ on $\mathcal{S}$ such that
\begin{equation}
 \sup_{A\subset\mathcal{S}} ||p(t,A;x_0)-\pi(A)||\le M(x_0)e^{-\alpha t},\ t\ge0,
\end{equation}
holds for some $\alpha>0$, all $x_0\in\mathcal{S}$ and some $M=M(x_0)>0$.
\end{theorem}
\blue{One of the main interests in this result lies in its similarity with commonly used Lyapunov conditions in systems and control theory \cite{Khalil:02}. In fact, the condition stated above is analogous to the convergence to a forward invariant set for deterministic dynamics, which also means that many of the available ideas and tools in the literature may be used in the current setting, with the caveat that the state-space is now discrete. However, this approach has the same drawbacks of Lyapunov methods -- that is, there is no general method for constructing such functions, but there is hope that the existing ideas could still be applied in the current context.

Linear Foster-Lyapunov functions were considered in by Gupta et al. \cite{Briat:13i}, who introduced the concept of ergodicity for the analysis of SRNs. The authors obtained various algebraic and computational conditions (e.g. linear or semidefinite programs) and connected them to analogous results in systems and control theory, especially to those arising in the analysis of (linear) positive systems \cite{Farina:00}. Interestingly, this approach showed that the circadian clock network of Section \ref{sec:NI:oscillations} and the stochastic switch of Section \ref{sec:NI:switching} are \textbf{structurally exponentially ergodic}, meaning that the exponential ergodicity property is independent of the rate parameters and is rather a topological property. Those results shed some light on the structure a network should possess in order to be (structurally) ergodic. Briat \& Khammash  \cite{Briat:20:Structural,Briat:19:DelayedRN} proposed extensions to address in a systematic way the robust/structural ergodicity of SRNs and SRNs with delays were proposed in. Milias-Argeitis \& Khammash \cite{Milias:14} considered more general polynomial Foster-Lyapunov functions, and Cappelletti \& Majumder \cite{Cappelletti:19b} addressed the ergodicity analysis of SRNs with stochastic time-varying rates, albeit using a different approach.}

\subsection{Moment boundedness and convergence}

\blue{In section~\ref{sec:models} we discussed several issues inherent to the moment equations but did not address the analysis problem. As already mentioned, the moment equations are (in most cases) open, which makes the application of standard approaches for the analysis of deterministic dynamical systems rather difficult. In particular, it renders the computation of equilibrium points, both analytically and numerically, infeasible.

Interestingly, ergodicity analysis provides an indirect way to properly approach this problem. Indeed, by virtue of the identity \eqref{eq:ensenble:erg}, the ergodicity test given in Theorem~\ref{th:ergodicity} can be used to prove the moments boundedness and their global convergence to their stationary values. Therefore, it can be used as a certificate for the existence of a unique globally asymptotically stable equilibrium point for the moment equations, at least up to a certain order. Other approaches to proving the boundedness and convergence of moments have also been developed in \cite{Engblom:12,Rathinam:13}.}

\blue{The approach described above corresponds to Lyapunov's direct method. Therefore, it seems important to explore analogues of Lyapunov's indirect method \cite{Khalil:02}. In this regard, some linearization methods for the local representation of the moment equations in the case of mass-action kinetics have been proposed. Until now the main underlying idea has been to first close the moment equation in order to obtain a nonlinear deterministic dynamical system, and then linearize it about its equilibrium points using standard methods; see e.g. \cite{Constantino:16,Vlysidis:18}. Unfortunately, this approach has severe drawbacks. The first is that it relies heavily on moment closure methods that are known to be inexact and for which there is no guarantee that the computed equilibrium points will be even close to the actual ones. The second is that it is tacitly assumed that equilibrium points exist without really proving their existence, which results in a circular argument. The ergodicity-based approach circumvents both issues while providing a more solid theoretical foundation by proving the existence of a globally asymptotically stable equilibrium point. Sensitivity results, such as those given by Gupta \& Khammash \cite[Theorem 3.2]{Gupta:14b}, can also be used to show continuity and even differentiability of the stationary moments with respect to the network parameters -- a necessary condition for proving the local validity of linearized models. In spite of those drawbacks, current linearization-based methods are very useful for design purposes such as controller design. We explain this further explained in Section \ref{sec:cybergenetics}.}

\section{Control of stochastic reaction networks}\label{sec:cybergenetics}

\blue{Having discussed the stability properties of SRNs, we are now in position to address the control of those networks. We briefly describe the different control problems that can be addressed in this context. We also describe the different constraints \emph{in-vivo} controllers need to satisfy to be implementable within living cells. We then discuss the possible implementations of integral controllers that can provably solve the regulation problem. Finally, we describe a few results on the \emph{in-vivo} and \emph{in-silico} control of single cells and cell populations.}

\subsection{Control problems in Cybergenetics}

\blue{The control problem in Cybergenetics is multifaceted, even when putting aside the control objectives that can be considered. The first option is choosing between whether we would like to control the behavior of an individual cell or that of a cell population; the second is choosing between whether the controller will be implemented inside a cell (\emph{in-vivo}) or a computer (\emph{in-silico}). In this section, we consider these options and focus on the \emph{robust perfect adaptation control problem}. At the single-cell level, this problem consists of finding a controller that can maintain the level of a molecule of interest $\Y$ (in a certain sense) to a desired set-point while, at the same time, rejecting constant external stimuli (i.e. disturbances). At the population level, the objective is to maintain the average level, across the cell population, of a molecule of interest $\Y$, again despite the presence of constant external stimuli.}

\blue{\emph{In-vivo} controllers are implemented within cells in terms of biological components and reactions, meaning that measurement, actuation, and all the other controller calculations are realized in terms of (noisy) chemical reactions. The external monitoring of the cells' behavior is usually achieved through the use of fluorescent proteins, which can be externally measured using time-lapse microscopy or flow cytometry. Whereas time-lapse microscopy allows for the in real-time tracking of the fluorescence levels within a single-cell or across a cell population, flow cytometry provides a snapshot of the current fluorescence distribution of the reporter proteins across the cell population at a given time. In this regard, flow-cytometry is more adapted to the observation of cell populations and their control.}

\blue{\emph{In silico} controllers, by contrast, are implemented inside a computer using traditional methods. Depending on the context, readouts from the cells are also obtained using either time-lapse microscopy or flow cytometry  \cite{Milias:11,Milias:16}. Actuation is performed using either light inputs (optogenetics) \cite{Rullan:18} or chemical induction \cite{Uhlendorf:12,Menolascina:14}. Interestingly, \emph{in-silico} control can also be used as a proxy for prototyping \emph{in-vivo} controllers by first implementing them \emph{in-silico} and then testing them on a single cell or a cell population \cite{Kumar:21,Kumar:22}.}

\blue{Finally, it is important to mention for completeness that control problems other than the robust perfect adaptation problem exist and have been considered. Examples include the tracking of more complex reference signals and the control of multi-cellular systems where the relative populations of two different cell types aims to be controlled; see \cite{Qian:18review} for further examples.}

\subsection{Implementation constraints in biological systems}\label{sec:constraints}

\blue{In this section, we review the different constraints that need to be satisfied by \emph{in-vivo} controllers.}

\blue{\subsubsection{High level of uncertainty}\label{sec:constraint:uncertainty}

A major difficulty in biological models lies in their inherent high level of uncertainty. The main underlying reason is that biology has only recently become a quantitative field, and for this reason, we still have a poor quantitative knowledge of how living organisms, let alone a single cell, fully work. As a result, our models will necessarily miss important elements, both at a molecular level (e.g. missing proteins) and at an interactional level (e.g missing (in)direct interactions between network components or poor quantifications of known reactions), leading to strong topological/structural, and parametric uncertainties. In this regard, if a controller circuit is to be implemented, it should be robust not only with respect to those types of model uncertainties but also with respect to its own uncertainties (i.e. resilient). A key to design circuits that satisfy such constraints is to design them in such a way that their function is a structural property that is also robust with respect to potential structural perturbations. This is the reason why \textbf{structural analysis and design} are playing, and will certainly play, an essential role in the analysis and design of biological circuits.}

\blue{\subsubsection{Noise}\label{sec:constraint:noise}

As already discussed, chemical reactions within cells are inherently stochastic and this may (and should) be exploited in the design of de-novo circuits. The presence of noise does not make the design of circuits easier but it may make the desired functions easier to achieve. A major difficulty is that the noise introduces some disorder in the execution of the circuit program by randomizing the execution order of the different calculations (i.e. the reactions) the circuit is performing. In this regard, those circuits should be robust with respect to their own random execution, which is a constraint that is absent from most usual engineering problems. While some of  the mechanisms behind noise-induced properties still need to be understood, it is commonly believed that they arise from the interplay between noise and nonlinear dynamics. Understanding them could allow us to design simpler circuits that naturally exploit noise to achieve their function.

There are currently no clear guidelines for the design of noise-exploiting circuits. Additional noise properties, such as attenuation or amplification, may also be useful and understanding how to design circuits exhibiting those properties is also fundamental. Once again there is no general way to design circuits with desired noise properties even though some mechanisms involved in signaling networks are already known.}

\subsubsection{Context dependence and cross-talk}\label{sec:constraint:dependence}

\bblue{Context dependence refers to the influence of the host cell on the execution of the implemented circuits.} Indeed, biological circuits all rely on shared resources (RNA polymerases, ribosomes, amino acids, ATP, etc.) managed by the cell, which invariably introduce a high level of coupling between all the circuits within the cells \cite{Qian:17ACS}. In this regard, it is essential to develop circuits that can mitigate the context dependence. They may also be designed so as to limit the use of the shared resources \cite{Frei:20,Jones:20} in order to avoid disrupting both the execution of the implemented circuits and the entire cell's operation. Closely related is the so-called \textbf{cross-talk} meaning that some molecules may be involved in multiple circuits, which may implement very different functions. In this regard, perturbing those molecules will interfere with all the circuits it is involved in, as well as all the other circuits those circuits are connected with, and so on.

\subsubsection{Energy Consumption and Mutational Escape}\label{sec:constraint:energy}

De-novo engineered circuits implemented in a host cell add burden or load to the host through the hijacking of energy and molecular resources, which may lead to cell death. In this regard, it is crucial to develop circuits with a low energy demand so as to limit the overall burden. One way to achieve this is to develop circuits that are small in size, work in a low-copy number regime, and are possibly noise-exploiting. \bblue{An interesting measure in the context of controllers is to compare the energy spent over time by a controller with that of a naive open-loop strategy, the latter having the lowest possible, but without all the closed-loop benefits.  Energy consumption can be reduced arbitrarily closely to that lower bound; see \cite{Briat:15e,Briat:16a}.  It is important to stress that while the energy cost of each reaction is rather difficult to quantify, this approach still allows to draw one interesting conclusions.}

Another related issue is that of \textbf{mutational escape} which is the phenomenon where a genetic mutation at a specific location in a given cell gives it an evolutionary advantage. This will ultimately lead to the domination of the progeny of that cell over the others as cells proliferate. If that mutation occurs on an implemented circuit and disrupts its function, then the overall cell population will, in the long-run, be devoid of functioning circuits. Strategies for overcoming this phenomenon or devising circuits that are evolutionary robust have started to appear \cite{Son:21}. A possible solution would be to link the circuit to an essential resource of the cell in such a way that the cell would die should the circuit stop functioning.

\subsubsection{Biorealizability}\label{sec:constraint:realization}

Functions to be implemented inside cells (e.g. mathematical functions or dynamic behaviors) need to be implemented in terms of elementary chemical reactions  (e.g. catalytic reactions) or modules (e.g. phosphorylation cycles, (incoherent) feedforward/feedback loops, antithetic/sequestering pairs, etc.).  A circuit/function that satisfies such constraints is called \textbf{biorealizable}. A necessary condition for biorealizability is that the function be representable as a positive system, because molecular counts and concentrations are nonnegative quantities.

Oishi \& Klavins \cite{Oishi:10b} notably introduced a procedure for turning any linear dynamical system into a chemical reaction network implementation based on the fact that any dynamical system can be expressed as the difference of two positive ones. Another approach is based on the use of regularization functions \cite{Briat:19:Logistic} which consists of forcing certain variables to be nonnegative. Finally, computational methods have been obtained for compiling static mathematical functions into elementary chemical reaction networks by exploiting the strong (uniform computability) Turing completeness of chemical reaction networks involving a finite number of molecular species (under the differential semantics) \cite{Fages:17,Hemery:21}. While these approaches may seem appealing for implementing any known system in terms of chemical reactions, they often lead to circuits that are too large and that do not exploit the inherent properties of reaction networks and other phenomena such as noise. Those are the main issues of such a top-down approach. In this regard, it seems much more promising to directly design the functions in the chemical reaction network domain, enabling one to incorporate the desired constraints and properties early on to design simpler, more efficient and reliable circuits. Finally, it is worth adding that all the existing results on the topic (with the exception of \cite{Fages:17}) are only for deterministic reaction networks, and that there is a clear need for analogous results in the stochastic setting where the impact of noise on biorealizability has remained elusive.\\

\subsection{Biorealization of integral controllers for robust perfect adaptation}

In this section, we discuss few reaction networks that are capable of integral action and their validity in the stochastic case. We also discuss an internal model principle for reaction networks exhibiting robust perfect adaptation.

\subsubsection{Absolute Concentration Robustness and Integral Control}\label{sec:ACR}

Absolute concentration robustness (ACR) is the property of a reaction network that the equilibrium values of some of the states are the same regardless of the value of the initial condition and/or certain rate parameters. Shinar \& Feinberg \cite{Shinar:10} obtained structural sufficient conditions within the framework of deficiency theory, which was developed independently by Horn \& Jackson \cite{Horn:72} and Feinberg \cite{Feinberg:72,Feinberg:95}. It was suspected for a long time that ACR was related to the presence of an integral action within the reaction network. This was proven to be true in by Cappelletti et al. \cite{Cappelletti:20}, who showed that the networks satisfying the conditions of Shinar \& Feinberg \cite{Shinar:10} necessarily embedded a certain class of nonlinear integral action, which Xiao \& Doyle \cite{Xiao:18b} referred to as  \emph{constrained integral controllers}. Karp \& al. \cite{Karp:12} and Dexter et al. \cite{Dexter:15} also considered more general invariants, but their connection to integral action is yet to be demonstrated.

The use of controller networks that make the closed-loop network exhibit ACR is a clever approach for solving the robust perfect adaptation in the deterministic setting \cite{Kim:20}. Unfortunately, Anderson \& al. \cite{Anderson:14,Anderson:17b} showed that the stochastic version of those networks irremediably has an absorbing state and will therefore exhibit a finite-time extinction; see Section \ref{sec:NI:extinction}.

\subsubsection{Bioregularization of integral control}

As mentioned above, some dynamical systems can be bioregularized to make them implementable in terms of chemical reactions. One example is the bioregularization of the standard integrator $\dot{I}=\mu-y$ using a \textit{positively regularizing function} $\varphi$ that satisfies certain properties (see \cite{Briat:19:Logistic} for more details). A simple one is $\varphi(I)=\exp(\alpha I)$, $\alpha>0$, and by defining $v:=\varphi(I)$ as a new state variable, we then obtain that $\dot{v}(t)=\alpha v(t)(\mu-y(t))$, which gives rise to the so-called \emph{autocatalytic integral controller} \cite{Drengstig:12}
\begin{equation}
  \V\rarrow{\alpha \mu}2\V,\ \V+\Y\rarrow{\alpha}\Y,\ \V\rarrow{k}\V+\X{1}
\end{equation}
where $\V$ is the controller species, $\Y$ is the controlled species, and $\X{1}$ is the actuated species of the SRN to be controlled. Other controllers, such as the logistic integral controller, can be obtained using such an approach (see \cite{Briat:16a,Briat:19:Logistic} for more details) and it turns out that all of those controllers belong to the abovementioned mentioned class of constrained integral controllers. While this approach is suitable for the design of deterministic controllers, it also fails to produce fully functional stochastic controllers due to the presence of at least one absorbing state at zero for the controller species, which is a common issue for all constrained positive integral controllers.

\subsubsection{Antithetic integral control}

The Antithetic Integral Controller (AIC) \cite{Briat:15e} is a biorealizable version of integral control based on the \emph{dual-rail representation} of the standard integrator \cite{Oishi:10b}. A possible reaction network implementation of this controller is given by
\begin{equation}
  \phib\rarrow{\mu}\Z{1},\ \Y\rarrow{\theta}\Yz+\Z{2},\ \Z{1}+\Z{2}\rarrow{\eta}\phib,\  \Z{1}\rarrow{k}\Z{1}+\X{1},
\end{equation}
where $\Z{1}$ and $\Z{2}$ are the controller species, $\Y$ is the measured/controlled species, and $\X{1}$ is the actuated species of the SRN to be controlled. The deterministic model of this controller is given by
\begin{equation}\label{eq:AIC:det}
    \dot{z}_1(t)=\mu-\eta z_1(t)z_2(t)\quad\textnormal{and}\quad \dot{z}_2(t)=\theta y(t)-\eta z_1(t)z_2(t).
\end{equation}
Letting $z:=z_1-z_2$, we obtain that $\dot{z}(t)=\mu-\theta y(t)$, showing that the difference between the two states behaves as an integrator with set-point $\mu/\theta$. The properties of the deterministic version of this nonlinear integrator have been studied by Briat \cite{Briat:19:Logistic}, Olsman et al. \cite{Olsman:19,Olsman:19b}, and Baetica et al. \cite{Baetica:20}. In the stochastic setting, the integral action acts at the level of the mean values for the state variables and the behavior of this network, while still achieving an integral action, is dramatically different. We explain this further in Section \ref{sec:invivo}.

The core mechanism is the sequestration reaction $\Z{1}+\Z{2}\rarrow{\eta}\phib$, which allows one to both perform a comparison between the two controller species (i.e. nonlinear difference) and effectively close the control loop. Sequestration is a well-known mechanism that has been quite well studied and several molecular mechanisms relying on or implementing it have been identified. Examples include toxin/anti-toxin \cite{Gerdes:88,DeJonge:09}, sigma-factors \cite{Chen:12,Annunziata:17,Aoki:19}, sense and anti-sense RNA \cite{Pelechano:13,Frei:22}, STAR/antisense RNA \cite{Lee:18,Agrawal:19}, sRNAS/mRNAs \cite{Kelly:18}, and scaffold/anti-scaffold \cite{Hsiao:15}, among others. Khammash and colleagues \cite{Filo:22,Frei:22} proposed monolithic extensions of this controller structure to incorporate proportional and derivative actions.

\subsubsection{Internal model principle for robust perfect adaptation}\label{sec:IMP}

A natural question is whether every reaction network exhibiting robust perfect adaptation involves an integral action. The answer to this question is affirmative and Gupta \& Khammash \cite{Gupta:22} proved that, in the case where the set-point is defined in terms of two reaction rates, perfect adaptation implies the presence of an integral action in the network dynamics.  In the deterministic setting, it was shown that the integral action can be implemented using at least one molecular species (in terms of constrained integral controllers), whereas in the stochastic setting, the integral action necessarily involves an antithetic pair. As a result, achieving robust perfect adaptation in the stochastic setting requires at least two molecules, which demonstrates that the AIC is the simplest integral controller that can be considered in that setting. This approach is essentially analytical; Araujo \& Liotta \cite{Araujo:18}  proposed a structural approach for establishing topological requirements for robust perfect adaptation.

%

\subsection{\emph{In-vivo} control for robust perfect adaptation}\label{sec:invivo}

\blue{In the stochastic setting, \emph{in-vivo} control can be used to achieve robust perfect adaptation at both the population and single-cell levels by exploiting the cells' noisy behavior and the fact that fluctuations will be averaged out when considering the cell population or the time-average at the single-cell level. Here, we discuss three control strategies: decentralized, centralized, and distributed control.}

\subsubsection{Decentralized population and single-cell control}

\blue{In a decentralized setup, each cell implements its own controller, which makes decision based only on local information, thereby ignoring the rest of the population. Briat at al. \cite{Briat:15e} proved that the AIC solves the robust perfect adaptation problem at the level of the average of a molecular species across the cell population provided the closed-loop network is ergodic. That is, the closed-loop network exhibits robust perfect adaptation and we have that
\begin{equation}
    \lim_{t\to\infty}\E[Y(t)]=\dfrac{\mu}{\theta}.
\end{equation}
Interestingly, the AIC exhibits noise-induced stability properties in the sense that while the deterministic closed-loop network may exhibit sustained oscillations, the average of the stochastic trajectories across the cell population globally converges to the desired steady-state values. This is illustrated in Figure~\ref{fig:stable}, which shows that the AIC achieves robust perfect adaptation for a simple gene expression network. The underlying mathematical and topological mechanisms behind this noise-induced property are still unclear, but as for all noise-induced properties, it emerges from the interplay between nonlinear dynamics and noise. However, this difference  with respect to the deterministic case can be interpreted by the presence of covariance terms in the corresponding moment equation
\begin{equation}
  \begin{array}{rcl}
    \dfrac{d}{dt}\E[Z_1(t)]&=&\mu-\eta \E[Z_1(t)]\E[Z_2(t)]-\eta \textnormal{Cov}[Z_1(t),Z_2(t)],\\
    \dfrac{d}{dt}\E[Z_2(t)]&=&\theta \E[Y(t)]-\eta \E[Z_1(t)]\E[Z_2(t)]-\eta \textnormal{Cov}[Z_1(t),Z_2(t)],
  \end{array}
\end{equation}
which are obviously absent from the deterministic model for the AIC given in \eqref{eq:AIC:det}. This induced property is of fundamental importance here because it makes the controller structurally functional independently of the values of its rate parameters. This is of crucial importance in synthetic biology as the accurate implementation of rate parameters is unfeasible.}

\blue{A drawback of the AIC controller is that it always increases the noise compared with a naive open-loop control strategy that would place the output at the desired level. One reason for this is that the controller is itself noisy and may combine its own noise with that of the system, thereby increasing the overall noise in the closed-loop network. Aoki et al. \cite{Aoki:19}  experimentally observed this theoretical prediction in the control of bacteria. The influence of a negative feedback loop was theoretically studied by Briat et al. \cite{Briat:18:Interface}, who showed that noise could be effectively reduced using negative feedback. This was later validated experimentally by Frei et al. \cite{Frei:22}, who showed that adding a proportional negative action effectively reduced the variability in the regulated molecular species. Olsman et al. \cite{Olsman:19,Olsman:19b}  extended the AIC in the deterministic setting to incorporate some quorum-sensing molecules to allow cells to communicate with each other and get information about the current population size. There is currently no analogue of this result in the stochastic case.}\\

\blue{Interestingly, by virtue of the ergodicity property of the closed-loop network, the same properties hold at the single-cell level where robust perfect adaptation is achieved in the sense that
\begin{equation}
  \lim_{t\to\infty}\dfrac{1}{t}\int_0^tY(t)\dt=\dfrac{\mu}{\theta}
\end{equation}
with probability one, which means that the long-term time-average of a single-cell trajectory also converges to the set-point $\mu/\theta$ (with probability one). As for population control, the noise at the single-cell level can be reduced through the use of a negative feedback loop.}\\

\subsubsection{Centralized population and single-cell control}

In the centralized control setup, all cells are controlled by a common controller or multiple instances of it. Such controllers could be present either in the environment/medium \cite{Duso:21} or within a different cell type that acts as controller cells and controls system cells \cite{Fiore:17}. The actuation and measurement reactions are implemented in this context using quorum-sensing molecules which can diffuse in and out of the cell's membrane and communicate with the external world.

Duso et al. \cite{Duso:21} addressed the centralized robust perfect adaptation problem using the AIC in the stochastic setting. The advantage of this approach is that it can keep track of the evolution of the population by considering cell divisions and cell deaths and can ensure perfect adaptation with respect to changes in the number of cells in the population. It can also be used to effectively control the number of cells within the population. These results are theoretical and still need to be validated experimentally.

\subsubsection{Distributed population and single-cell control}

\blue{In the distributed context, the controllers are located within every cell in the population but also communicate with each other. Olsman et al.  \cite{Olsman:19,Olsman:19b} considered the case of communicating cells, but the case of communicating controllers has not been addressed in either the deterministic or the stochastic setting. This problem remains completely open.}

\subsection{\emph{In-silico} control for robust perfect adaptation}\label{sec:insilico}

\blue{\emph{In-silico} control can also be used to achieve robust perfect adaptation at both the single-cell and population levels. We discuss the same control strategies in this context.}

\subsubsection{Decentralized population and single-cell control}

Kumar and colleagues \cite{Kumar:21,Kumar:22} reported decentralized \emph{in-silico} control of single cells and cell populations, where each cell is allocated a specific controller. This work has only been experimental so far and no theory has been developed in this context in either the deterministic or stochastic setting. However, it is worth mentioning that certain results obtained for the decentralized \emph{in-vivo} control may be applied to this setup. More recently, Briat \& Khammash \cite{Briat:21:Optimal}  theoretically addressed the optimal and $H_\infty$ control of stochastic reaction networks using Dynamic Programming. The controllers were characterized in terms of the solution of non-standard Riccati differential and difference equations for which a new theory has been developed.

An interesting fact is that such an approach yields closed-loop networks that contain both continuous-deterministic and discrete-stochastic dynamics whenever deterministic controllers are used to control stochastic reaction networks, or vice-versa. This gives rise to a closed-loop network expressed as a Piecewise-Deterministic Markov Process (PDMP) \cite{Davis:93} for which no general theory has been developed in this specific scenario. Another interesting fact is that mean-field CTMCs and PDMPs can also be obtained when the control input of the SRNs is computed from the moments of the molecular species of that SRN (for an analogous setup in power networks, see e.g., \cite{Meyn:15}). Here also, the problem is mostly open and no theory has been developed to address this problem in this particular context.

\subsubsection{Centralized population and single-cell control}

Briat \& Khammash \cite{Briat:12c,Briat:13h,Briat:19:Opto} theoretically addressed the centralized control of noisy cell populations in the moment equations framework. Reference \cite{Briat:12c} first showed that the stationary mean and variance of the controlled species could be independently assigned within their admissibility region by considering two independent control inputs. Experiments reported by by Gerhardt et al \cite{Gerhardt:21} also led to the same conclusions, albeit using a different approach. This result needs to be contrasted with \emph{in-vivo} control where it is possible to act on the mean and the variance of the controlled species independently but using one single control input \cite{Briat:18:Interface,Frei:22}.

Briat \& Khammash \cite{Briat:13h,Briat:19:Opto} later extended the moment to consider a more complex network exhibiting the moment closure problem. By combining ergodicity and sensitivity analysis, they developed a linearization approach that avoided the use of closure methods. The main difficulty here stemmed from the fact that the unknown higher-order moments had to be considered unknown inputs that should be rejected at stationarity by the integral controller. The fact that the exact value of those higher-order moments was unknown led to the need to consider a continuum of equilibrium points, which all needed to be made stable in closed-loop. While this approach was successful for that specific system, it is unclear whether it can be extended to more complex networks.

\subsubsection{Distributed population and single-cell control}

The \emph{in-silico} distributed control of cell populations is a problem that has not really been addressed so far and is therefore fully open. In this setup, the controllers are allowed to share information with each other following the same communication topology as cells use to communicate with each other. In this context, sharing information among the different controllers is much simpler than by using \emph{in-vivo} control thanks to the fact that controllers are implemented \emph{in-silico}. This is mostly an open problem that has only been partially addressed so far in an experimental setting (see e.g. \cite{Chait:17}). The theoretical aspect is still completely open.

\subsection{Filtering of stochastic reaction networks}

It seems important to briefly talk about the filtering problem in this context due to its potential role in the implementation of certain types of controllers. Filtering is an essential problem in signal processing and control theory where the goal is to design a system that will be able to denoise a given signal or estimate unmeasurable variables despite the presence of noise or disturbances. Well-known filters include those based on the Kushner-Stratonovich and the Zakai equations and the Kalman-Bucy filter (see e.g. \cite{Bain:09}). However, filtering is also present at a biochemical level in circuits that are trying to attenuate noise or to anticipate a certain phenomenon; e.g. in signaling pathways. Until now the design of \emph{in-vivo} filters that can estimate unmeasured molecular species (but also rate parameters) within a single cell has been quite limited and the only work on the topic was by Zechner at al. \cite{Zechner:16}, whose approach is based on a reaction network implementation of the Kushner-Stratonovich equation \cite{Bain:09} specialized to the case of CTMCs. Such a filter is known to be nonlinear and finite-dimensional and to reduce to the well-known Kalman-Bucy filter in the linear Gaussian case. The filter was validated \emph{in-vitro} but remains to be implemented and validated in a living organism.

On the other hand, Khammash and colleagues \cite{Milias:16,Fang:21,Fang:21b}  have developed \emph{in-silico} (particle) filters in both the deterministic and the stochastic settings. There is hope that if such filters can be implemented in real-time, they will play a large role in the \emph{in-silico} control of single-cells and cell populations. Indeed, all of the approaches that have been developed for the control of living organisms at a biomolecular level have assumed that the measured output is also the controlled output. Filters have the potential to relax this strong assumption and to allow for the control of molecular species that are not directly measured. This problem has not been addressed so far.

\section{Discussion}


In this section we make some reflections related to the topics discussed and the path forward.



\subsection{Models complexity and reliability}

\blue{When compared with models in other fields, biological models tend to be fairly inaccurate and often fail to provide meaningful predictions for experimental outcomes. This is largely due to the sheer complexity of biological phenomena and the difficulty in measuring biological variables
owing to the small scale of biological components and their general inaccessibility. This exposes a critical limitation, as the true power of mathematical models resides in their ability to make accurate predictions. In a biological setting, the development of accurate predictive models  would lead to a dramatic improvement in efficiency, as experiments could finally be guided by simulations before being attempted, similar to what is already done in engineering and other fields. While detailed whole-cell models are beginning to emerge \cite{Karr:12,Goldberg:18}, a lot of advancements will be needed before such models are as accurate as their counterparts in many fields of engineering.}


This dichotomy between biological models and the reality they aim to describe is too often forgotten, especially in heavily model-based disciplines such as control theory, where models are just accurate enough for their purpose. This large gap between models and reality is also present in other fields, such as in economics/finance, ecology and epidemiology. Not unlike the situation in biology, the underlying reason for this gap lies in the complex nature of the considered systems, which involve a large number of constituents and which evolve on many different temporal and spatial scales.  Even if relatively more accurate mechanistic models can be painstakingly developed, they are often large and complex, resulting in poor scalability of the overall modeling paradigm. Model simplifications are then considered for tractability, at the expense of a loss of accuracy. One natural question then, is how to obtain models that are more reliable while keeping the complexity at a reasonable level.

\blue{This question has multiple possible answers depending on the objective considered. When the goal is modeling reality, models will need to yield accurate simulations and predictive results. When our objective is controlling a reaction network, we need a model that captures the features of the network that contain information about stability and control-relevant performance. The main issue with mechanistic models is their high complexity, their potential lack of identifiability, and possibly their lack of accuracy, especially when working with stochastic dynamics. This suggests that there is a need for radically new ideas but, as also pointed out by Vitttadello \& Stumpf \cite{Vittadello:22}, it is also possible that our current knowledge of mathematics is simply not ready for such a challenge.}

In spite of that, we will suggest two alternative approaches to pure mechanistic modeling. The first is to use phenomenological models that do not necessarily have a precise physical meaning but may still accurately capture enough of the input/output behavior of a system by combining simple dynamical elements in a clever way. Similarly, the combination of mechanistic and phenomenological components could also be considered in order to keep the complexity low while still maintaining some physical meaning (see e.g. \cite{Olson:14}). Such models have been considered  with great success, and it would be interesting to clarify the source of their accuracy. In this regard, it seems important to characterize the different reliable possible phenomenological components that can be used and to understand in which situations they could be used. Until now, there has not seemed to be any solid theory about such a phenomenological approach for the modeling of complex systems. Developing such a theory could greatly simplify the models of such systems while preserving their accuracy, thereby providing an effective solution to this important question.

The second option would be to consider data-based methods in order to completely avoid using models, and instead rely exclusively on data. These data may then serve as a basis for making predictions, designing circuits, or even establishing certain properties such as ergodicity. However, data-based methods still lack the support of researchers in communities where more reliable and trustable models exist. While this may be true in many fields, it is far from being the case, for complex systems. For example, in the context of cybergenetics, possible applications could be the data-driven optogenetic control of single-cells and cell populations based on model-free methods.

\subsection{Analysis of SRNs}

\blue{We have presented some analysis results based on the ergodicity of CTMCs. While Gupta et al. \cite{Briat:13i} provided some insights on the topological constraints that may or may not lead to ergodicity properties, their work did not include all possible network structures. A large amount of work is still needed in order to provide better ergodicity certificates and to get a better understanding of the topological properties that lead to ergodic behaviors.}

\blue{From a more systems and control theory perspective, the analysis of interconnections of reaction networks could be addressed in a similar way as in robust analysis and control (e.g. through use of multipliers or scalings) in order to be able to consider larger networks, possibly at a reduced computational cost \cite{Briat:14d}.  However, while interconnections in systems theory are usually information flow exchanges, interconnections in biology may also consist of mass exchanges. As with electric circuits, the loading effect (the fact that a network perturbs an upstream network when it connects to it) may lead to a loss of the desired behavior upon interconnection. This motivated the design of biological insulators that could be placed between circuits in order to preserve the desired behavior and achieve modularity (see e.g. \cite{DelVecchio:15b}). In this regard, the analysis of interconnections exhibiting a loading effect needs to be addressed using specific tools that are still lacking. Finally, it is important to introduce performance measures and develop accompanying computational methods to evaluate properties other than robust perfect adaptation.}

\subsection{Linearization and control of the moment equations}

\blue{The moment equations, although useful, suffer from the lack of closure. Until now there has been no real effort to quantify the error, even at stationarity, of proposed closure methods. Having such bounds would allow us to linearize them and take into account the degree of uncertainty using robust analysis and control methods. Alternative linearization methods based on ergodicity and sensitivity analysis may also be of interest in order to generalize the ideas described by Briat \& Khammash \cite{Briat:19:Opto}.

The control of the moment equations may also take a completely new turn using data-driven methods. Recall that the moment closure problem is a model-based problem that is devoid of any data-based representation of the system. In this regard, it seems quite appealing to explore the possibility of addressing the centralized control of cell populations using a data-based representation of the system. Note that the data will contain all the necessary information about the higher-order moments,  and therefore, unlike the moment equation, the data will be ``closed". An alternative method would be to close the moment equations using real-world data from the process. Ruess et al. \cite{Ruess:11} considered similar ideas using synthetic SSA data.}

\subsection{Control theory for systems and synthetic biology}

\blue{Section \ref{sec:cybergenetics} described several important theoretical robust perfect adaptation problems that remain open. However, it is also important to mention that other types of reference inputs may be interesting to track (e.g. oscillatory signals). This has been addressed only at the experimental level, and no theory seems to have been developed in this context. This problem is highly related to the internal model principle, which remains to be developed in the context of stochastic reaction networks. A simple question is, Would considering two AICs in series be enough to achieve the tracking of a ramp signal in the stochastic setting?}\\

\blue{We now briefly discuss closed-loop networks represented as PDMPs. While some theoretical work has been developed for such systems \cite{Costa:08}, it remains sporadic in the fields of systems biology and, a fortiori, control theory for systems biology (see e.g. \cite{Li:17b}). In this regard, deriving tractable ergodicity conditions for such systems would be of great interest and preliminary steps have been made in this direction (see e.g.\cite{Benaim:18} and the references therein). The main difficulty here is showing that the state-space is irreducible for the dynamics of the system due to the hybrid nature of the state space and, in particular, because of its continuous component.}

\subsection{Random programs}

As mentioned in Section \ref{sec:constraint:noise}, noise randomizes the execution of the implemented genetic program. Here, each reaction can be seen as an instruction or a line of code that is executed randomly according to the propensity function associated with the reaction/instruction. This is to be contrasted with sequential execution, which is the way controller programs are executed in computers. Interestingly, the underlying reason why such programs work is that the randomness (i.e. the reaction rates) is actually state-dependent. It seems quite clear that if those rates were all made either constant or state-independent, many of the properties would no longer hold, at least not structurally. In this regard, it seems interesting to formalize those programs, understand how they execute and clarify what structure they need to possess to be well-behaved (e.g. terminate, converge, achieve their goal, etc.). A potential way of doing that is through the consideration of stochastic Petri nets \cite{Haas:02} which have been already considered for the modeling of deterministic and stochastic reaction networks in \cite{Heiner:13,Liu:14b,Baez:17}. However, they have never been considered for the stability and performance analysis of (controlled) reaction networks, and this mostly remains an open problem.

\subsection{Reaction networks design}

\blue{We have argued throughout this article that bottom-up approaches should be favored in order to fully exploit biological mechanisms when designing \emph{de-novo} circuits. In particular, we may ask whether modularity, as mentioned earlier in this section, is achievable or even desirable in a biological context. Modularity is a fundamental, flexible engineering paradigm that allows us to design complex systems from the interconnection of elementary building blocks. This successful principle has been, for instance, at the core of electrical engineering, industrial, and manufacturing systems design. However, many natural complex systems do not seem to have been built based upon this principle -- possibly because those complex systems developed through an evolutionary, selective process. Indeed, the presence of cross-talk and shared resources within cells results in a high coupling within cells and a lack of clear modular structure.}

\blue{In light of this, it is not clear whether the design of efficient synthetic biological circuits should necessarily be achieved through the interconnection of elementary building blocks following a strictly modular approach or should rather be designed in a holistic way in order to fully exploit all biological phenomena (nonlinearities, noise, etc.) and respect all cellular constraints (low metabolic cost, etc.) An example is the class of monolithic AIC-based PID controllers developed by Filo et al. \cite{Filo:21b,Alexis:22} and Alexis et al. \cite{Alexis:22} that do not rely on the interconnection of independent proportional, integral and derivative actions, but rather exploit the inherent possibilities of different interactions around the AIC. This non-modular design is, in fact, much less complex than a modular one would have been.}

\blue{In Section \ref{sec:constraints} we gave reasons to explain why implementing circuits that work in a low-copy-number regime could be beneficial. In particular, noise-driven designs were expected to be capable of leading to simpler, more efficient designs than is possible in the deterministic setting. Yet, the identification of all of the topological mechanisms (in terms of the reaction network topology) behind noise exploitation, suppression, and amplification remains elusive. Other mechanisms that remain to be understood are those that are noise-susceptible (i.e. fail with noise) or noise-resilient (i.e. function with noise). Understanding those properties at a topological level would provide us with invaluable guidance in the design of noise-exploiting circuits. Finally, it is also important to identify functions or behaviors that cannot be achieved in the presence of noise.}

\subsection{Beyond Cybergenetics}

\blue{While this article has focused exclusively on reaction networks and their control in a biological setting, it is also interesting to explore whether those analysis tools and control structures may be useful in a different setting, such as in engineering. Perhaps more importantly, it would be compelling to find examples of engineering problems where noise-induced properties arise and could be exploited. Moreover, as reaction networks essentially describe population models, this suggests that the tools and ideas discussed in this article will also apply to population models and may therefore be useful in fields such as ecology and epidemiology.}

\section*{Acknowledgments}

The authors are grateful to Timothy Frei, Maurice Filo, Ankit Gupta, Andreas Milias-Argeitis, Noah Olsman, and Christoph Zechner for their comments on a preliminary version of this article.


\end{document}